\documentclass[article]{revtex4-1}
\usepackage{epstopdf}
\usepackage[T1]{fontenc}
\usepackage[utf8]{inputenc}
\usepackage[english]{babel}
\usepackage{amsmath,amssymb,amsfonts}
\usepackage{graphicx,graphics}
\usepackage{subfig}
 \usepackage{caption}
 \usepackage{dcolumn}
\usepackage{bm}
\usepackage{color}
\usepackage{amsthm}             
\usepackage{amssymb}
\usepackage{mathrsfs}
\usepackage{dcolumn}
\usepackage{bm}
\usepackage{epsfig}
\usepackage{epstopdf}
\usepackage{float}
\begin{document}

\title{ Control of atomic transition rates via laser light shaping}

\author{R. J\'auregui}
\affiliation{Instituto de F\'{\i}sica, Universidad Nacional Aut\'onoma de M\'exico}
\address{Apartado Postal 20-364, 01000, M\'exico D.F., M\'exico.}
\email{rocio@fisica.unam.mx}
\begin{abstract}
A  modular systematic analysis of the feasibility of modifying atomic transition rates by tailoring the electromagnetic field of an external coherent light source is presented.
The formalism considers both the center of mass and internal degrees of freedom of the atom, and all properties of the field: frequency, angular spectrum, and polarization.  General features of recoil effects for internal forbidden transitions are discussed. A comparative analysis of different  structured light sources  is explicitly worked out. It includes spherical waves, Gaussian beams,  Laguerre-Gaussian beams, and propagation invariant beams with closed analytical expressions.  It is shown that increments in the order of magnitude of the transition rates for Gaussian and Laguerre-Gaussian beams, with respect to those obtained in the paraxial limit, requires waists of the order of the wavelength, while propagation invariant modes  may considerably enhance transition rates under more favorable conditions.  For  transitions that can be naturally described as modifications of the atomic angular momentum, this enhancement  is maximal (within propagation invariant beams) for Bessel modes, Mathieu modes can be used to entangle the internal and center of mass involved states, and Weber beams  suppress this kind of transitions unless they have a significant component of odd modes. However, if a recoil effect of the transition with an  adequate symmetry is allowed,  the global transition rate (center of mass and internal motion) can also be enhanced using Weber modes. The global analysis presented reinforces the idea that a better control of the transitions between internal atomic states requires both a proper control of the available states of the atomic center of mass, and shaping of the background electromagnetic field.
\end{abstract}
\maketitle
\section{Introduction.}
Modifications of the internal states of an atom due to its coupling to the electromagnetic (EM) field are, under standard conditions, highly restricted. Once the proper selection rules are satisfied, the enhancement or inhibition of transition rates can be based on the control of the evolution of the atomic center of mass, and the appropriate selection of the features of the surrounding electromagnetic field. The latter includes its frequency spectrum, angular spectrum and polarization. In this work, we make a systematic study of the feasibility of modifying atomic transition rates by tailoring the electromagnetic field, given different  schemes for the center of mass dynamics.

In general, previous analysis that had the enhancement or inhibition of internal state transition rates in mind, can be classified as those which study atoms confined in electromagnetic cavities \cite{cavidades} or nearby surfaces \cite{chew},
and studies in which laser light is shaped in open space to create electromagnetic fields with a predetermined
structure \cite{nienhuis,Davila,jauregui,alexandrescu,klimov}. The first scenario, in general, gives rise to polarization  restrictions: the natural modes would be, {\it e. g.}, transverse electric or transverse magnetic EM fields satisfying the adequate boundary conditions on the cavity surface. In the last scenario, most descriptions usually assume circular or linear polarization with respect to the main propagation axis of a given laser beam. Here we present a general analysis that includes both cases.

We  take into account the relevance of the central field description of electrons in atoms, and
the fact that EM fields can  be described as superpositions of modes with well defined total (orbital plus polarization) angular momentum.
The accompanying symmetries of both atom and EM field are essential to get a proper description of the
transition rates under realistic conditions. In all the calculations, the role of the atomic center of mass will be  emphasized.

In Section II, a general accurate description of the so called scalar and vectorial EM modes via their wave vector spectra is presented; using it guarantees the validity of our forthcoming analysis beyond the limitations of the usual paraxial approximation. In Section III,  the transition matrix elements of the minimal coupling interaction Hamiltonian of a nonrelativistic hydrogen like atom  are written in terms of products of a center of mass factor, an internal atomic state factor and an electromagnetic factor which couples the former two terms. Based on this, a modular description of transition rates involving all the degrees of freedom of the system atom-EM field is developed. In Section IV, we revisit the general properties of the internal transition factor and the standard classification in terms of the  coupling of different atomic multipoles to the EM field. We also show that the developed formalism can be used to study internal transitions which require a relativistic description without modifications on the center of mass and electromagnetic factors. In Section V we stress the main features of the center of mass term when it is described by a quantum wave function. The limitations of such an approach under realistic conditions are also briefly discussed. In order to compare with standard studies of atomic recoil effects, in Section VI, we apply our formalism when the center of mass states is described in terms of plane waves. In Section VII, we report the explicit calculation of the electromagnetic factor for spherical modes, Gaussian modes and propagation invariant modes for a center of mass motion with spherical symmetry. This could correspond, for instance, to atoms in free space or in a spherical trap. In section VIII, an optimization procedure to enhance transition rates by selecting the proper parameters of a Laguerre-Gaussian beam or of a Bessel beam is explicitly worked out; the trap that confines the cold atoms is assumed to have circular cylinder symmetry. Finally, some conclusions derived from this study are given in Section IX.

\section{ The light field: structured beams and their  angular spectra.}
 The electric field $\mathbf{E}(\mathbf{r},t)$ of an electromagnetic  wave in otherwise free space can always  be written as a superposition of plane waves with wave vector $\mathbf{k}$, frequency $\omega$ and polarization $b$:
 \begin{equation}
\mathbf{E}(\mathbf{r},t) =\sum_b\sum_{\kappa}\int d\omega \mathcal{S}(\omega)\int d^3\mathbf{k} e^{i(\mathbf{k}\cdot\mathbf{r} -\omega t)}\delta (\vert k_z \vert -\sqrt{\omega^2/c^2 - k_\bot^2}) {\mathfrak{E}}^{(b)}_{\kappa}(\mbox{k}_x ,\mbox{k}_y;\omega).\label{eq:EF}\end{equation}
In this equation $\mathcal{S}(\omega)$ denotes the frequency spectra, the delta function guarantees the fulfillment of the dispersion relation,  $k_\bot =\sqrt{k_x^2 +k_y^2}$, and
${\mathfrak{E}}^{(b)}_{\kappa}(\mbox{k}_x ,\mbox{k}_y;\omega)$ is the  vectorial spectrum in $\mathbf{k}$ space for the polarization $b$; it is characterized by a set of parameters $\kappa$ that are required to specify  each electromagnetic mode of frequency $\omega$.
The magnetic field associated to the EM wave is given by
\begin{equation}
\mathbf{B}(\mathbf{r},t) =\sum_b\sum_{\kappa}\int d\omega \mathcal{S}(\omega)\int d^3\mathbf{k} e^{i(\mathbf{k}\cdot\mathbf{r} -\omega t)}\delta (\vert k_z\vert -\sqrt{\omega^2/c^2 - k_\bot^2})
 {\mathfrak{B}}^{(b)}_{\kappa}(\mbox{k}_x ,\mbox{k}_y;\omega)\end{equation}
with
\begin{eqnarray}
{\mathfrak{B}}^{(b)}_{\kappa}(k_x ,k_y;\omega)&=& (c/\omega) {\mathbf{k}}\times {\mathfrak{E}}^{(b)}_{\kappa}(k_x ,k_y;\omega).
\end{eqnarray}

In the Coulomb gauge, the electromagnetic vector potential for each mode is
\begin{eqnarray}
\mathbf{A}^{(b)}_{\kappa}({\bf{r}},t) &=& \int d^3\mathbf{k} e^{i(\mathbf{k}\cdot\mathbf{r} -\omega t)}\tilde{\mathbf{A}}^{(b)}_{\kappa}({\bf{k}};\omega)\\
&=&\frac{ic}{\omega} \int d^3\mathbf{k} e^{i(\mathbf{k}\cdot\mathbf{r} -\omega t)}\delta (\vert k_z \vert -\sqrt{\omega^2/c^2 - k_\bot^2}){\mathfrak{E}}^{(b)}_{\kappa}(k_x,k_y;\omega).\label{eq:A}
\end{eqnarray}
 The polarization $b$  reflects the transverse structure of the electric field $\nabla \cdot   \mathbf{E}(\mathbf{r},t)=0$, so that,
\begin{equation}{\mathbf{k}}\cdot{\mathfrak{E}}^{(b)}_{\kappa} =0.\end{equation}
One selection of the resulting two independent polarizations is defined in terms of a preselected vector $\mathbf{a}$ as follows (from now on, $\hat{\mathbf{v}}$ denotes the unitary operator along $\mathbf{v}$):
\begin{eqnarray}
{\mathfrak{E}}^{TE}_{\kappa}(k_x ,k_y;\omega)&=& \mathcal{E}_\kappa^{TE}[\hat{\mathbf{k}}\times {\mathbf{a}}]
\tilde{\Psi}^{(TE)}_{\kappa;\omega}(k_x ,k_y),\label{eq:bmodeE}\\
{\mathfrak{E}}^{TM}_{\kappa}(k_x ,k_y;\omega)&=& \mathcal{E}_\kappa^{TM}[\hat{\mathbf{k}}(\hat{\mathbf{k}} \cdot {\mathbf{a}})- {\mathbf{a}}]\tilde{\Psi}^{(TM)}_{\kappa;\omega}(k_x , k_y).\label{eq:bmodeM}
\end{eqnarray}
These modes are the well known  transverse electric ($TE$) and transverse magnetic ($TM$) modes frequently used in cavity electrodynamics (although they can also be generated in free space \cite{kj}).
The vector ${\mathbf{a}}$ in such systems is usually chosen as an axis of symmetry of the cavity.
The normalized functions $\tilde{\Psi}^{(b)}_{\kappa;\omega}(\mbox{k}_x ,\mbox{k}_y)$  are known in the literature as the angular spectra of the mode, the factor
$\mathcal{E}^{(b)}$ is proportional to the mode amplitude. Specific examples of $\tilde{\Psi}^{(b)}_{\kappa;\omega}(\mbox{k}_x ,\mbox{k}_y)$ are given below.
If ${\mathbf{a}}$ is parallel to the $z$ axis, the beams are  said to be vectorial EM structured beams with a privileged propagation axis along the $z$-axis; with this selection
\begin{equation}
\hat{\mathbf{a}}=\hat{\mathbf{e}}_z\equiv\hat{\mathbf{e}}_0 \quad  {\mathrm{and}} \quad \hat{\mathbf{e}}_\pm = ({\mathbf{e}}_x \pm i {\mathbf{e}}_y)/\sqrt{2}, \label{eq:pol}
\end{equation}
and
\begin{eqnarray}
\mathbf{e}^{(TE)}\equiv \hat{\mathbf{k}}\times \hat{\mathbf{e}}_z = &\frac{k_\bot c}{\sqrt{2}\omega i}(e^{i\varphi_{\mathbf{k}}}\hat{\mathbf{e}}_- - e^{-i\varphi_{\mathbf{k}}}\hat{\mathbf{e}}_+)&  \nonumber\\
\mathbf{e}^{(TM)}\equiv \hat{\mathbf{k}}(\hat{\mathbf{e}}_z \cdot \hat{\mathbf{k}})- \hat{\mathbf{e}}_z =  &\frac{k_\bot k_z c^2}{\sqrt{2}\omega^2 }(e^{i\varphi_{\mathbf{k}}}\hat{\mathbf{e}}_- + e^{-i\varphi_{\mathbf{k}}}\hat{\mathbf{e}}_+)& - \frac{k_\bot^2 c^2}{\omega^2} \hat{\mathbf{e}}_z . \label{eq:polarizations}
\end{eqnarray}
Here $\varphi_{\mathbf{k}}$ is the azimuthal angle in the wave vector space, $e^{\pm i\varphi_{\mathbf{k}}} = (k_x \pm i k_y)/k_\bot$.

Another selection of polarization basis originates from the paraxial description of light beams. In this case,
the so called scalar structured beams are described as follows. First, a fixed direction is selected $\hat{\mathbf{a}}_3$, and two other orthogonal directions are identified $\hat{\mathbf{a}}_2,$ and
$\hat{\mathbf{a}}_1$. Then,
\begin{eqnarray}
{\mathfrak{E}}^{1}_{\kappa}(k_1 ,k_2;\omega)&=& \mathcal{E}_\kappa^{1}[\hat{\mathbf{a}}_1 - \frac{\mathbf{k}\cdot \hat{\mathbf{a}}_1}{\mathbf{k}\cdot \hat{\mathbf{a}}_3}\hat{\mathbf{a}}_3]\tilde{\Psi}^{(1)}_{\kappa;\omega}(k_1 ,k_2)
\equiv \mathcal{E}_\kappa^{1}\mathbf{e}^{(1)}\tilde{\Psi}^{(1)}_{\kappa;\omega}(k_1 ,k_2),\label{eq:smode1}\\
{\mathfrak{E}}^{2}_{\kappa;\omega}(k_1 ,k_2;\omega)&=& \mathcal{E}_\kappa^{2}[\hat{\mathbf{a}}_2 - \frac{\mathbf{k}\cdot \hat{\mathbf{a}}_2}{\mathbf{k}\cdot \hat{\mathbf{a}}_3}\hat{\mathbf{a}}_3]\tilde{\Psi}^{(2)}_{\kappa;\omega}(k_1 ,k_2)\equiv\mathcal{E}_\kappa^{2}\mathbf{e}^{(2)}\tilde{\Psi}^{(2)}_{\kappa;\omega}(k_1 ,k_2),\label{eq:smode2}
\end{eqnarray}
In the paraxial limit  $\hat{\mathbf{a}}_3$ usually  corresponds to the main direction of propagation of the beam. To be specific, let it be the $z$-axis, $i.$ $e.$, $\hat{\mathbf{a}}_3=\hat{\mathbf{e}}_z$. The paraxial condition $\vert k_z\vert \sim \omega/c$ yields $ ({\mathbf{k}\cdot \hat{\mathbf{a}}_i})/(\mathbf{k}\cdot \hat{\mathbf{a}}_3) \ll 1$, $i=$1, 2, so that the component of the field along $\hat{\mathbf{a}}_3$ is usually neglected; if  $\hat{\mathbf{a}}_1=\hat{\mathbf{e}}_x$ and $\hat{\mathbf{a}}_2=\hat{\mathbf{e}}_y$, $\mathbf{a}_1$ and $\mathbf{a}_2$ define the two orthogonal linear polarization directions.

Scalar EM modes may, under adequate circumstances, be approximately generated using a monochromatic laser beam with the proper polarization that impinges a space light modulator or a structured diffraction plate. To that end, each optical device is respectively programmed or designed to guarantee the appropriate boundary condition that defines each mode at the plane $z=0$:
\begin{eqnarray}
\hat{\mathbf{a}}_i\cdot\mathbf{E}_{\kappa,\omega}(x,y,0) &=&\sum_b\int d^3\mathbf{k} e^{i(k_x x +k_y y )}\delta (\vert k_z \vert -\sqrt{\omega^2/c^2 - k_\bot^2})\hat{ \mathbf{a}}_i\cdot{\mathfrak{E}}^{(b)}_{\kappa}(\mbox{k}_x ,\mbox{k}_y;\omega)\nonumber\\
&=&\int_{\mathcal{C}_{\omega/c}} d^2\mathbf{k} e^{i(k_x x +k_y y )}\mathcal{E}_\kappa^{i}\tilde{\Psi}^{(i)}_{\kappa;\omega}(k_x ,k_y)\label{eq:circ}\\
\Leftrightarrow
\mathcal{E}_\kappa^{i}\tilde{\Psi}^{(i)}_{\kappa;\omega}(k_x ,k_y) &=&\frac{1}{(2\pi)^2}\int dx\int dy \quad e^{-i(k_x x +k_y y )}\hat{\mathbf{a}}_i\cdot\mathbf{E}_{\kappa,\omega}(x,y,0)
.\label{eq:generate}\end{eqnarray}
In Eq.~(\ref{eq:circ}), $\mathcal{C}_{\omega/c}$ denotes the interior surface of a circumference of radius $\omega/c$ in the $\mathbf{k}_\bot$ space.

Vectorial EM modes can be written (an generated) as superpositions of scalar EM modes and viceversa by the selection of the proper angular spectra;
taking $\mathbf{a} = \hat{\mathbf{a}}_3$,
 \begin{eqnarray}
 \mathbf{e}^{(TE)}=\hat{\mathbf{k}}\times \hat{\mathbf{a}}_3 &=&\quad
 \hat{k}_2[\hat{\mathbf{a}}_1 - (\hat{k}_1/\hat{k}_3) \hat{\mathbf{a}}_3] -\quad\hat{k}_1[\hat{\mathbf{a}}_2 - (\hat{k}_2/\hat{k}_3)\hat{\mathbf{a}}_3]
 =\quad\hat{k}_2\mathbf{e}^{(1)} - \quad\hat{k}_1\mathbf{e}^{(2)},\nonumber\\
 \mathbf{e}^{(TM)}=\hat{\mathbf{k}}(\hat{\mathbf{a}_3} \cdot \hat{\mathbf{k}})- \hat{\mathbf{a}}_3&=&
 \hat{k}_1\hat{k}_3[\hat{\mathbf{a}}_1 - (\hat{k}_1/\hat{k}_3)\hat{\mathbf{a}}_3]+
 \hat{k}_2\hat{k}_3[\hat{\mathbf{a}}_2 - (\hat{k}_2/\hat{k}_3)\hat{\mathbf{a}}_3] = \hat{k}_1\hat{k}_3\mathbf{e}^{(1)}+ \hat{k}_2\hat{k}_3 \mathbf{e}^{(2)}.
 \end{eqnarray}
 Notice that the superpositions depend on $\mathbf{k}$.
   Quantization of structured modes in terms of vectorial beams \cite{shj} is usually more direct than with scalar modes due to the evident orthogonality between $TE$ and $TM$ modes, in contrast to the  overlap between the scalar modes defined by Eqs.~(\ref{eq:smode1}-\ref{eq:smode2}).

The purpose of this work is to present a general formalism to compare, in a quantitative way, the dependence of atomic transition amplitudes for different selections of polarization and angular spectra  of a coherent light source. For the selection $\mathbf{a} = \hat{\mathbf{e}}_z = \hat{\mathbf{a}}_3$, explicit results will be given for the following angular spectra :
\begin{itemize}
\item{} Gauss modes with waists $\mathrm{w}_x$ and $\mathrm{w}_y$:
\begin{equation}\tilde{\Psi}^{(b)}_{\kappa_\bot,m;\omega}(k_x ,k_y) =\mathcal{N}_{\mathrm{w}_x,\mathrm{w}_y;\omega}^{(b)} e^{-\mathrm{w}_x^2k_x^2}e^{-\mathrm{w}_y^2k_y^2}\label{def:gauss:as}\end{equation}

\item{} Laguerre Gaussian modes with waist $w_o$, topological charge $m$ and order $p$:
\begin{equation}\tilde{\Psi}^{(b)}_{w_o,m,p;\omega}(k_x ,k_y) =\mathcal{N}_{\mathrm{w}_o,m,p;\omega}^{(b)} e^{im\varphi_{\mathbf{k}_\bot}} \sum_{a =0}^p
(-1)^a\begin{pmatrix} p + m \\ p-a\end{pmatrix} 2^{a+m/2 -1}\Big(\frac{k_\bot w_0}{c}\Big)^m e^{-k_\bot^2w^2_o/4}L_a^m\Big(k_\bot^2w^2_o/4\Big).\label{def:laggauss:as}\end{equation}

\item{} Propagation invariant beams with analytical expressions. These beams, under idealized conditions, maintain its transverse structure with respect to a given direction that here we take as $\hat{ \mathbf{e}}_z$. They have the angular spectra structure
    \begin{equation}\tilde{\Psi}^{(b)}_{\kappa_\bot,\kappa_m;\omega}(k_x ,k_y) =\mathcal{N}_{\kappa_\bot,\kappa_m;\omega}^{(b)} \frac{\delta(k_\bot - \kappa_\bot)}{k_\bot} \psi^{\kappa_m}(\varphi_{\mathbf{k}}).\label{def:piof:as}\end{equation} Explicitly:
\begin{itemize}
\item{} In circular cylinder coordinates, they are known as Bessel beams \cite{bessel,shj}: \begin{equation}\tilde{\Psi}^{(b)}_{\kappa_\bot,m;\omega}(k_x ,k_y) =\mathcal{N}_{\kappa_\bot;\omega}^{(b)}  \frac{\delta(k_\bot - \kappa_\bot)}{k_\bot} e^{im\varphi_{\mathbf{k}}}\label{def:bessel:as}\end{equation}

\item{} In elliptic coordinates with an ellipticity parameter  $q$ and order $m$, they are known as Mathieu beams \cite{mathieu,brlrjm}:  \begin{equation}\tilde{\Psi}^{(b)}_{\kappa_\bot,m;\omega}(k_x ,k_y) =\mathcal{N}_{\kappa_\bot,m;\omega}^{(b)} \frac{\delta(k_\bot - \kappa_\bot)}{k_\bot} (ce_m(\varphi_{\mathbf{k}};q) +i se_m(\varphi_{\mathbf{k}};q))\label{def:mathieu:as}.\end{equation}
where $ce_m(\varphi;q)$ and $se_m(\varphi;q)$ are Mathieu functions of order $m$ in standard notation \cite{nist}.
\item{} In parabolic coordinates, they are known as Weber beams \cite{weber,brlrjw}. For even parity and  aperture $\alpha$: \begin{equation}\tilde{\Psi}^{(b)}_{\kappa_\bot,\alpha;\omega}(k_x ,k_y) =\mathcal{N}_{\kappa_\bot,\alpha;\omega}^{(b)} \frac{\delta(k_\bot - \kappa_\bot)}{k_\bot} \frac{e^{i\alpha \ln\vert\tan\varphi_{\mathbf{k}}/2\vert}}{2\sqrt{\pi\vert\sin\varphi_{\mathbf{k}}\vert}}\label{def:weber:as}.\end{equation}
    Odd parity modes are obtained by replacing $\vert\sin\varphi_{\mathbf{k}}\vert$ in the denominator by $-i\vert\sin\varphi_{\mathbf{k}}\vert$.
\end{itemize}
\end{itemize}
Each of these elementary modes can be combined to yield, for instance, even standing transverse Bessel modes $\tilde{\Psi}^{(b)}_{\kappa_\bot,m;\omega}(k_x ,k_y)+\tilde{\Psi}^{(b)}_{\kappa_\bot,-m;\omega}(k_x ,k_y)$ or traveling Weber modes by
superposing even and odd modes with a $\pm\pi/2$ phase factor. The experimental generation of approximate propagation invariant beams can be implemented by
replacing the delta factor by a Gaussian function of $k_\bot$, centered around $\kappa_\bot$, and with a width much smaller than $\kappa_\bot$.

The calculation of the normalization factors $\mathcal{N}_{\kappa;\omega}^{(b)}$ is discussed in Appendix A. Their selection is done to allow an equitable comparison between
the effects of different modes in the atomic transition rates.

We shall also study the results obtained for vectorial spherical waves. In this case, the vector $\mathbf{a}$ that defines the polarizations is chosen as the gradient operator in $\mathbf{k}$ space, $\mathbf{\nabla}_{\mathbf{k}}$ and it is applied to the spherical harmonics in that space. Spherical electromagnetic waves are described in detail in Appendix B.

\section{Atom-Light interaction.}
\subsection{Coupling of the electron current to the electromagnetic field: the $H_{I1}$ interaction Hamiltonian.}
For simplicity we consider a hydrogen like atom in conditions that do not require a relativistic treatment.
In the next section, we describe how these restrictions can be relaxed to include multi electron atoms and relativistic effects.

We are interested in the evaluation of the transition probability by which an ultra cold atom,
initially described by a wave function that  can be written as a superposition
of functions that are products of a center of mass function $ \Phi_{\kappa^{CM}}({\bf R})$
and an internal function $\phi_{\kappa^{rel}}({\bf r}_{rel})$ with quantum numbers ${\kappa^{CM}}$ and
${\kappa^{rel}}$ respectively,
$$\Psi_0(\mathbf{R},\mathbf{r}_{rel}) = \Phi_{\kappa^{CM}_0}({\bf R})\phi_{\kappa^{rel}_0}({\bf r}_{rel}),$$
 changes its state to a wave function with the same structure, that is a superposition of wave functions of the form
$$\Psi_F(\mathbf{R},\mathbf{r}_{rel}) =\Phi_{\kappa^{CM}_F}({\bf R})\phi_{\kappa^{rel}_F}({\bf r}_{rel}),$$ and simultaneously absorbs
a photon described by a mode of the EM field ${\bf A}^{(b)}_{\bf\kappa}$, with polarization $b$ and structure labeled by $\kappa$.
In the former expressions the position of the electron is given by the coordinate \begin{equation}\mathbf{r}_e = \mathbf{R} +\frac{\mu}{M_e}\mathbf{r}_{rel},\label{eq:re}\end{equation} where $\mathbf{R}$ refers to the atom's center of mass, $M_e$ is the electron mass, $\mu=M_eM_N/(M_e + M_N)$ is the reduced mass, $M_N$ is the nucleus mass,  and $\mathbf{r}_{rel}$ is the relative coordinate. $\Phi_{\kappa^{CM}}({\bf R})$ and $\phi_{\kappa^{rel}}({\bf r_{rel}})$ are eigenfunctions of the corresponding Schr\"odinger equation with initial eigenvalues $E^{(0)}_{CM}$ and $E^{(0)}_{rel}$, and final eigenvalues $E^{(F)}_{CM}$ and $E^{(F)}_{rel}$.

We focus -- mostly-- in the case that the transition is induced by the interaction Hamiltonian
\begin{equation}
\hat H_{I1} = - \frac{q_e}{M_e} {\bf p}_{rel}\cdot {\bf A}_{\bf \kappa}^{(b)}({\bf r}_e), \label{eq:hi1}
\end{equation}
 with $q_e$, the electron charge and $\mathbf{p}_{rel}$ the relative momentum. This Hamiltonian can be understood as a result of the coupling
of the electron current  $q_e{\mathbf{p}_e/M_e}$ to the electromagnetic field ${\bf A}_{\bf \kappa}^{(b)}({\bf r}_e)$ in the approximation
where $\mathbf{p}_e = (M_e/(M_e+M_N)) \mathbf{P} + \mathbf{p}_{rel}\sim \mathbf{p}_{rel}$ valid whenever
$M_N\gg M_e$ ( which is satisfied under standard conditions). Corrections to this approximation are described below.

If ${\bf A}_{\bf\kappa}^{(b)}$ satisfies Coulomb gauge ( $\nabla\cdot{\bf A}_{\bf\kappa}^{(b)}=0$)
within first order perturbation theory, the transition amplitude $\mathrm{T_{0F}}$
 between the elementary wave functions $\Psi_0(\mathbf{R},\mathbf{r}_{rel})$ and $\Psi_F(\mathbf{R},\mathbf{r}_{rel})$  is given by
\begin{eqnarray}
\mathrm{T_{0F}} =\langle\Psi_F\vert \hat H_{I1}\vert\Psi_0\rangle &=&\frac{1}{i\hbar}(E^{(0)}_{rel} - E^{(F)}_{rel})\int d^3r_{rel}d^3R
\big[\Psi^*_F({\bf r}_{rel},{\bf R}){\bf r}_{rel}\Psi_0({\bf r}_{rel},{\bf
R})\big]\nonumber \\
&\cdot& \big[q_e\frac{\mu}{M_e}{\bf A}_{\bf \kappa}^{(b)}({\bf
R}+\frac{\mu}{M_e}{\bf r}_{rel}) \big],\label{eq:trans}
\end{eqnarray}
 in the length form.
Taking into account Eq.~(\ref{eq:A}),
\begin{equation}
{\bf A}_{\bf \kappa}^{(b)}({\bf r}_e) =
=\frac{ic}{\omega} \int d^3\mathbf{k} e^{i(\mathbf{k}\cdot (\mathbf{R} +\frac{\mu}{M_e}\mathbf{r}_{rel}) -\omega t)}\delta (\vert k_z \vert -\sqrt{\omega^2/c^2 - k_\bot^2}){\mathfrak{E}}^{(b)}_{\kappa}(k_x,k_y;\omega).
\end{equation}
 The structure of Eqs.~(\ref{eq:bmodeE}-\ref{eq:bmodeM}) and Eqs.~(\ref{eq:smode1}-\ref{eq:smode2}) let us write the   absorption transition amplitudes $\mathrm{T_{0F}}$  in terms of the light angular spectrum $\tilde{\Psi}^{(b)}_{\kappa;\omega}$ as follows
\begin{eqnarray}
\mathrm{T_{0F}}
 = \eta_{rel}\frac{c}{\omega}\int d\omega S(\omega) e^{-i\Delta t}
 \int d^3&k& \delta(k_z -\sqrt{\frac{\omega^2}{c^2} - k_\bot^2})
 I_{CM}(\mathbf{k};\kappa^{CM}_0;\kappa^{CM}_F)
 \mathbf{J}_{rel}(\mathbf{k};\kappa^{rel}_0;\kappa^{rel}_F)\cdot {\mathbf{e}_\kappa^{(b)}} \tilde{\Psi}^{(b)}_{\kappa;\omega}(k_x,k_y)
\label{eq:TOF}  \\
I_{CM}(\mathbf{k};\kappa^{CM}_0;\kappa^{CM}_F)&=& \int d^3R\Phi^*_{\kappa^{CM}_0}({\bf R}) e^{i\mathbf{k}\cdot R}\Phi_{\kappa^{CM}_F}({\bf R}) \label{eq:ICM} \\
 \mathbf{J}^{+}_{rel}(\mathbf{k};\kappa^{rel}_0;\kappa^{rel}_F)&=&  \int d^3\mathbf{r}_{rel} \phi^*_{\kappa^{rel}_0}(\mathbf{r}_{rel})e^{+i\frac{\mu}{M_e}\mathbf{k}\cdot \mathbf{r}_{rel}}{\bf r}_{rel}\phi_{\kappa^{rel}_F}(\mathbf{r}_{rel}) \label{eq:Jrel}\\
 \eta_{rel} = \frac{q_e \mu\omega_{rel}}{M_e},\quad\omega_{rel} &=& \frac{E^{(F)}_{rel} - E^{(O)}_{rel}}{\hbar},\quad\omega_{CM} = \frac{E^{(F)}_{CM} - E^{(0)}_{CM}}{\hbar},
\quad \Delta = \omega -\omega_{rel}-\omega_{CM}\label{eq:para}.
\end{eqnarray}
The corresponding emission amplitudes induced by the presence of the coherent EM field $\mathrm{T}_{F0}$ satisfy the relation $\mathrm{T}_{F0}=\mathrm{T}_{0F}^*$.
Eq.~(\ref{eq:para}) decomposes the transition amplitude $\mathrm{T}_{0F}$ in factors that involve separately the internal degrees of freedom of the atom $\mathbf{J}_{rel}$, and the degrees of freedom of the center of mass  $I_{CM}$. These factors couple, in a modular form, to a coherent EM field via its angular spectra $\tilde{\Psi}^{(b)}_{\kappa;\omega}(k_x,k_y)$, and its polarization ${\mathbf{e}_\kappa^{(b)}}$ (which in general depends on $\mathbf{k}$ in a way determined by $\kappa$ and $b$).

 \subsection{ Other interaction terms resulting from the nonrelativistic minimal coupling Hamiltonian.}

 The non relativistic Hamiltonian of two charged particles (the nucleus and the electron) interacting with the light electromagnetic field via the minimal coupling scheme, involves, besides the Hamiltonian $\hat H_{I1}$, two terms.
  One of them, for a monochromatic wave at resonance $\Delta =0$, gives rise to the transition amplitude
   \begin{eqnarray}
\langle\Psi_F\vert \hat H_{I2}\vert \Psi_0\rangle &=&
i\omega_{CM}\int d^3r_{rel}d^3R
\big[\Psi^*_F({\bf r_{rel}},{\bf R}){\bf R}\Psi^*_0({\bf r}_{rel},{\bf
R})\big] \big[q_e{\bf A}_{\bf \kappa}^{(i)*}({\bf
R}+\frac{\mu}{M_e}{\bf r}_{rel}) +q_N{\bf A}^{(i)*}_{\bf \kappa}({\bf
R}-\frac{\mu}{M_N}{\bf r}_{rel})\big]\nonumber
\\
&-&i\frac{q_N\mu}{M_N}\omega_{rel}\int d^3r_{rel}d^3R
\big[\Psi^*_F({\bf r}_{rel},{\bf R}){\bf r}_{rel}\Psi^*_0({\bf r}_{rel},{\bf
R})\big] {\bf A}_{\bf
\kappa}^{(i)*}({\bf R}-\frac{\mu}{M_N}{\bf
r}_{rel}).\label{eq:HI2}
\end{eqnarray}
with $q_N$ the  nucleus charge and $\mathbf{r}_N = \mathbf{R} -(\mu/M_N)\mathbf{r}_{rel}$ its position. The first term, in the last equation, results from the difference between the electron momentum and the momentum of the relative coordinate in the electron current-EM field coupling, the second and the third represent  the coupling between the nucleus current
 $q_N{\mathbf{p}_N/M_N}$ to the electromagnetic field ${\bf A}_{\bf \kappa}^{(b)}({\bf r}_N)$.

 For atoms trapped in a harmonic potential with characteristic frequency $\Omega$, the center of mass motion can be naturally measured in terms of $R\sim \sqrt{\hbar/\Omega M_T}$ while the transition frequencies are integer multiples of $\omega_{CM}\sim \Omega$, thus $\omega_{CM} R \sim \sqrt{\hbar\Omega/ M_T}$
    which  corresponds to $\sim 0.1 m/seg$ for a harmonic trap with $\Omega\sim 100 KHz$ and a hydrogen atom. For the internal motion $\omega_{rel} \sim Z^2\alpha_{fsc}^2 \mu c^2/2\hbar$ and $r_{rel}\sim \hbar/Z\mu c\alpha_{fsc}$, with $\alpha_{fsc}$ the fine structure constant; so that $\omega_{rel} r_{rel} \sim Z \alpha_{fsc} c/2$. Thus, $$q_e\omega_{rel} r_{rel}/q_N\omega_{CM} R \sim \alpha_{fsc} c\sqrt{M_T/\hbar\Omega} \sim 10^7\gg 1.$$ That is, the natural units to measure the matrix element
$$ \omega_{rel} q_e\langle\phi_{\kappa_F^{CM}}\vert  e^{i\mathbf{k}\cdot\mathbf{R}}\vert \phi_{\kappa_0^{CM}}\rangle
 \langle\phi_{\kappa_F^{rel}}\vert \mathbf{r}_{rel}e^{i\mathbf{k}\cdot\mathbf{r}_{rel}}\vert \phi_{\kappa_0^{rel}}\rangle,$$
     that defines the transition amplitude $\mathrm{T}_{OF}$, are much bigger than those related to
 $$ \omega_{CM} q_N\langle\phi_{\kappa_F^{CM}}\vert \mathbf{R} e^{i\mathbf{k}\cdot\mathbf{R}}\vert \phi_{\kappa_0^{CM}}\rangle
 \langle\phi_{\kappa_F^{rel}}\vert e^{\pm i\mathbf{k}\cdot\mathbf{r}_{rel}}\vert \phi_{\kappa_0^{rel}}\rangle.$$
       Since, under standard conditions $\mu\ll M_T $ and $q_N\ge q_e$, the order of magnitude of the relevant transition rates induced by
        any term in $\hat H_{I2}$ is expected to be much smaller than its equivalent one induced by $\hat H_{I1}$.

  Note, however, that transition amplitudes of $\hat H_{I2}$, Eq.~(\ref{eq:HI2}), can also be written in a modular form similar to Eq.~(\ref{eq:para}) that involves the vector integral
   \begin{eqnarray}
   \mathbf{I}_{CM}(\mathbf{k};\kappa^{CM}_0;\kappa^{CM}_F)&=& \int d^3R\Phi^*_{\kappa^{CM}_0}({\bf R}) e^{i\mathbf{k}\cdot \mathbf{ R}}\mathbf{R}\Phi_{\kappa^{CM}_F}({\bf R}), \label{eq:ICM2} \\
   &=& -i\mathbf{\nabla}_{\mathbf{k}}I_{CM}(\mathbf{k};\kappa^{CM}_0;\kappa^{CM}_F)
   \end{eqnarray}
   and the scalar integrals
   \begin{eqnarray}
   J^{\pm}_{rel}(\mathbf{k};\kappa^{rel}_0;\kappa^{rel}_F)&=&  \int d^3\mathbf{r}_{rel} \phi^*_{\kappa^{rel}_0}(\mathbf{r}_{rel})e^{\pm i\frac{\mu}{M_e}\mathbf{k}\cdot \mathbf{r}_{rel}}\phi_{\kappa^{rel}_F}(\mathbf{r}_{rel}), \label{eq:Jrel1}
   \end{eqnarray}
   for the first two summands in Eq.~(\ref{eq:HI2}), and the scalar integral  $I_{CM}$, Eq.~(\ref{eq:ICM}), and the vector integral
   \begin{eqnarray}
 \mathbf{J}^{-}_{rel}(\mathbf{k};\kappa^{rel}_0;\kappa^{rel}_F)&=&  \int d^3\mathbf{r}_{rel} \phi^*_{\kappa^{rel}_0}(\mathbf{r}_{rel})e^{-i\frac{\mu}{M_e}\mathbf{k}\cdot \mathbf{r}_{rel}}\mathbf{r}_{rel}\phi_{\kappa^{rel}_F}(\mathbf{r}_{rel})= \mathbf{J}^+_{rel}(-\mathbf{k};\kappa^{rel}_0;\kappa^{rel}_F)  \label{eq:Jrel2}\\
   &=& -i\mathbf{\nabla}_{\mathbf{k}}J^{-}_{rel}(\mathbf{k};\kappa^{rel}_0;\kappa^{rel}_F)
    \end{eqnarray}
   for the third summand in Eq.~(\ref{eq:HI2}). So that the main features of the modular evaluation of the transition amplitudes developed in the next pages
   can also be applied to $H_{I2}$.

 The third term $\hat H_{I3}$ in the minimal coupling Hamiltonian
 \begin{equation}
 \hat H_{I3} =  \frac{q_e^2}{2M_e} \vert \hat{\bf A}({\bf r}_e)\vert^2 +\frac{q_N^2}{2M_N} \vert \hat{\bf A}({\bf r}_N)\vert^2 \label{eq:HI3}
 \end{equation}
 is usually interpreted in terms of  two photon processes: introducing  a complete set of intermediate center of mass and relative coordinate states the corresponding transition amplitudes are $(q_{e,N}/2M_{e,N})\sum_I\langle\Psi_F\vert \mathbf{A}_\kappa^{(b)}(\mathbf{r}_{e,N})\vert \Psi_I\rangle\cdot\langle\Psi_I\vert \mathbf{A}_\kappa^{(b)}(\mathbf{r}_{e,N})\vert \Psi_0\rangle$. These transition amplitudes are proportional to $\alpha_{fsc}^2$ and, thus, generally smaller than the relevant ones arising from the $\hat H_{I1}$ Hamiltonian. Note that the $\hat H_{I3}$ matrix elements can also be written in a modular form involving the scalar integrals $J^{\pm}_{rel}$ and $I_{CM}$ defined above.

 It is important to mention that working with the minimal coupling scheme in the Coulomb gauge  gives equivalent  results \cite{cohen} to those obtained in the  PZW formalism \cite{PZW, lembessis}. This formalism  is obtained by applying a gauge transformation to a minimal coupling Lagrangian. It makes evident the relevance of the
 magnetic and electric dipole moments of the neutral atom evaluated with respect to the atomic center of mass; the electric and the magnetic fields of the photon field appear directly in the resulting interaction Hamiltonian which also includes a term quadratic in the EM field. Detailed studies of transition probabilities of atomic systems have been performed  using  the linear electric field part of the interaction Hamiltonian of the PZW formalism for a Laguerre-Gaussian beam in the paraxial approximation \cite{Davila,alexandrescu}.

\section{ Internal state factor $\mathbf{J}_{rel}$.}

Notice that  the  matrix element $ \mathbf{J}^+_{rel}$, Eq.~(\ref{eq:Jrel}), contains the information of the final and initial internal states of the atom and corresponds to that worked out in standard atomic physics calculations. From it the matrix element $ \mathbf{J}^-_{rel}$, Eq.~(\ref{eq:Jrel2}), can be calculated.
In this section we revisit key features of these matrix elements that are essential to follow the forthcoming calculations.

 Whenever the internal state of an atom is properly described by a central field interaction, that is, if $\psi_{\kappa_{rel}}({\bf r}) = {R}_{nlm}(r)Y_{lm}(\hat{\mathbf{r}})$  (with $Y_{lm}$ the $l$, $m$ spherical harmonic) the internal state quantum numbers are $\kappa_{rel} = n$, $l$, $m$,  and a relevant relation that makes  easier
 the evaluation of the relative coordinate integrals $\mathbf{J}_{rel}$  is
 \begin{equation}
 e^{i{\mathbf{k}}\cdot{\mathbf{r}}} = 4\pi \sum_{\ell,m_{\ell}} i^{\ell} j_{\ell}(kr)Y_{\ell,m_\ell}(\hat{\mathbf{k}})Y_{\ell,m_\ell}^*(\hat{\mathbf{r}}),
\label{eq:sph} \end{equation}
 which expresses plane waves in terms of spherical waves, in this equation $j_{\ell}$ is the spherical Bessel function of order $\ell$.  Each term in this series is an eigenfunction of the orbital
 angular momentum operators, $\hat{L}^2 = (\bf{r}\times \bf{\nabla})\cdot(\bf{r}\times \bf{\nabla})$ and $\hat{L}_z = (\bf{r}\times \bf{\nabla})_z$ .
 If $Y_{\ell,m_\ell}(\hat{\mathbf{k}})$ were taken as an angular spectra and the proper polarization vectors were used, vectorial spherical  electromagnetic modes which carry orbital angular momentum would be obtained  as described in Appendix B.
 Using Eq.~(\ref{eq:sph}), it results
  \begin{eqnarray}
   \mathbf{J}^{+}_{rel}(\mathbf{k};\kappa^{rel}_0;\kappa^{rel}_F) &=& \sum_{\ell,m_\ell,m_{ph}}
\mathfrak{J}^{\ell,m_\ell,m_{ph}}(\mu\omega/c M_e;\kappa^{rel}_o;\kappa^{rel}_F)\hat{\mathbf{e}}^*_{m_{ph}}Y_{\ell,m_\ell}(\hat{\mathbf{k}}),\nonumber \\
\mathfrak{J}^{\ell,m_\ell,m_{ph}}(\mu\omega/c M_e;\kappa^{rel}_o;\kappa^{rel}_F)&=&\mathfrak{R}^{(1)}_{\ell,m_\ell,\kappa^{rel}_o,\kappa^{rel}_F}(\mu\omega/c M_e)
\upsilon^{(4)}(l,l^\prime,\ell,1;m_o,m_F,m_\ell,m_{ph}), \label{eq:Jplus}
 \end{eqnarray}
with $\hat{\mathbf{e}}_{-m_{ph}}$, $m_{ph} =0,\pm 1$ the polarization vectors defined in Eqs.~(\ref{eq:pol}),
$$\mathfrak{R}^{(i)}_{\ell,m_\ell,,\kappa^{rel}_o,\kappa^{rel}_F}(\mu\omega/cM_e) = \int_0^\infty dr r^{2+i}{R}^*_{n_ol_om_o}(r) j_{\ell}(\mu\omega r/M_ec){R}_{n_F l_F m_F }(r),  $$
and the angular integral
$$\upsilon^{(4)}(l_o,l_F,\ell,1;m_o,m_F,m_\ell,m_{ph})= 4 \pi i^\ell \int d\Omega_{\mathbf{r}}Y_{\ell,m_\ell}^*(\hat{\mathbf{r}})Y^*_{l_om_o}(\hat{\mathbf{r}})Y_{l_F m_F}(\hat{\mathbf{r}})Y_{1m_{ph}}(\hat{\mathbf{r}}).$$
The dependence of $\mathbf{J}^+_{rel}(\mathbf{k};\kappa^{rel}_0;\kappa^{rel}_F)$  on $\mathbf{k}$ is through its modulus $\omega/c$ and
its orientation $Y_{\ell,m_\ell}(\hat{\mathbf{k}})$ as a manifestation of the Wigner-Eckart theorem.

Particular useful forms of ${R}_{nlm}(r)$ correspond to linear superpositions of Slater type orbitals (STO) $\mathrm{e}^{-\beta r} r^{n-l-1}$. For them, the relevant integral can be directly computed
$$Q_{\beta_o + \beta_F;\omega}^{\ell,STO}=\int_0^\infty dr e^{-(\beta_o + \beta_F)r} r^s j_\ell(\mu\omega r/M_ec)
=\frac{\sqrt{\pi}}{2^{\ell +1}(\beta_o +\beta_F)^{s+1}}\frac{\Gamma(\ell + s +1)}{\Gamma(\ell +3/2)}$$
\begin{equation}
\Big(\frac{\mu\omega}{cM_e(\beta_o +\beta_F)}\Big)^\ell \quad_2\mathrm{F}_1\Big(\frac{\ell+s+1}{2},\frac{\ell+s+2}{2},\ell +3/2,-\Big(\frac{\mu\omega}{M_ec(\beta_o + \beta_F)}\Big)^2\Big).
\end{equation}
For a hydrogenic atom $\beta = (Z/n)(\alpha_{fsc}\mu c^2/\hbar c)$. Since the energy associated to changes in the center of mass state of an atom involve in general energies much less than those of electronic states, close to resonance $\omega\sim \omega_{rel}\sim (Z^2\alpha_{fsc}^2\mu c^2/2\hbar)(1/n_o^2 - 1/n_F^2)$ and the term
$$\frac{\mu\omega}{M_ec(\beta_o + \beta_F)}\sim \frac{Z\alpha_{fsc}\mu}{2M_e}\Big(\frac{1}{n_o} - \frac{1}{n_F}\Big).$$
Note that, unless $Z\alpha_{fsc}$ is not too small, $i.$ $e.$, for  atoms requiring a relativistic description of their internal state, the series that define $_2\mathrm{F}_1$ can be cut at the zeroth order term $_2\mathrm{F}_1\sim 1$. That is,  under non relativistic conditions, retardation effects of the atom-light interaction can be neglected, and the internal atom transitions in terms of static electric and magnetic multipoles is valid. Besides, for $Z\alpha_{fsc} \ll 1$, $Q_{\beta_o + \beta_F;\omega}^{\ell,STO}$ rapidly decreases as $\ell$ increases: the highest possible values of $\mathfrak{J}^{\ell,m_\ell,m_{ph}}(\mu\omega/c M_e;\kappa^{rel}_o;\kappa^{rel}_F)$ can be found for $\ell =0$, that is,  electric dipole transitions induced by the Hamiltonian $\hat H_{I1}$ . For such transitions, the companion spherical harmonic $Y_{00}$ does not depend on $\hat{\mathbf{k}}$. The next order term in the series of $\mathbf{J}^+_{rel}(\mathbf{k};\kappa^{rel}_0;\kappa^{rel}_F)$ depends on $\hat{\mathbf{k}}$ through $Y_{1,m}(\hat{\mathbf{k}})$. If the quantization axis of the atom coincides with the $z$-axis,
 \begin{equation}Y_{1,0}(\hat{\mathbf{k}}) =\sqrt{\frac{3}{4\pi}} \frac{{k}_z c}{\omega},\quad \quad Y_{1,\pm 1}(\hat{\mathbf{k}}) = \sqrt{\frac{3}{8\pi}}\frac{{k}_\bot c}{\omega} e^{\pm i\varphi_\mathbf{k}}.\label{eq:Y1m}\end{equation}
This term can be related to electric quadrupole transitions and magnetic dipole transitions  induced by the Hamiltonian $\hat H_{I1}$. However the latter would not be properly described unless electron spin effects had been incorporated.

The angular structure of $\mathfrak{J}^{\ell,m_\ell,m_{ph}}$ given by $\upsilon^{(4)}$ reflects the vectorial character of the electromagnetic
field through the presence of the  $Y_{1,-m_{ph}}$ factor, and the relevance of angular structure of that field through the orbital angular momentum term $Y_{\ell,m_\ell}$. Both of them can be coupled to yield an effective total angular momentum
\begin{equation}
Y_{1,-m_{ph}}(\hat{\mathbf{r}})Y_{\ell,m_\ell}(\hat{\mathbf{r}})=
\sum_{\ell_T,m_T}\sqrt{\frac{(2\ell+1)(3)(2\ell_T+1)}{4\pi}}\begin{pmatrix}1&\ell&l_T\\-m_{ph}&m_\ell&m_T\end{pmatrix}\begin{pmatrix}1&\ell&l_T\\0&0&0\end{pmatrix}
Y_{l_Tm_T}(\hat{\mathbf{r}})
\end{equation}
A similar expression can be used for $Y^*_{l_0,m_0}Y_{l_F,m_F}$ so that the angular integrals $\upsilon^{(4)}$ have a closed expression in terms of the 3-$j$ symbols. The selection rules for each $\ell$ multipole result from the structure of 3-$j$ symbols.

 For the evaluation of the transition amplitudes related to the Hamiltonians $\hat H_{I2}$, Eq.(\ref{eq:HI2}), and  $\hat H_{I3}$, Eq.~(\ref{eq:HI3}), it is necessary to calculate the  internal scalar transition factors ${J}^{\pm}_{rel}$, Eq.~(\ref{eq:Jrel1}). Using the  series expansion given by Eq.~(\ref{eq:sph}), these factors  can also be written in a completely analogous form to that of the vector factors $\mathbf{J}^{\pm}_{rel}$. Now  the functions $\mathfrak{R}^{(0)}_{\ell,m_\ell,\kappa^{rel}_o,\kappa^{rel}_F}$ and $\upsilon^{(4)}(l_o,l_F,\ell,0;m_o,m_F,m_\ell,0)$ are the relevant ones. Whenever retardation effects can be neglected the standard condition $kr_{rel}<<1$ makes that the  term $\ell =0$ in the series expansion of the scalar factor $J^{\pm}_{rel}$ gives a negligible value, and the terms with lowest $\ell$ value that may give a significant contribution correspond to $\ell =1$ , $m_\ell =0,\pm 1$ instead of the $\ell =0$ term of the electric dipole transitions induced by $\hat H_{I1}$. That is,  the scalar integral ${J}^{\pm}$ term associated to  dipole transitions has the structure ${J}^{\pm}_{dipole,rel}\sim\mathfrak{R}^{(0)}\upsilon^{(4)} Y_{1,m_\ell}(\hat{\mathbf{k}})$. A similar argument follows for higher multipoles.

For  multiple electrons atoms, most of the equations given above are still valid both in the context of the independent particle model with a central field approximation, and for treatments that yield natural orbitals \cite{natorb} after including electronic correlation effects.

A relativistic description of the internal state evolution of a hydrogen like atom can be done in terms of the solution of the Dirac equation. Then the effective interaction Hamiltonian for the electron-EM field coupling could be taken as $\hat H_D = \gamma_\mu A^\mu$, whenever the CM motion is not relativistic, with $\gamma_\mu$ the Dirac matrices and $A^\mu$ the electromagnetic potential evaluated at the position of the electron undergoing the atomic internal transition, Eq.~(\ref{eq:re}). The plane wave exponential would then be written as a product of an exponential term dependent on the CM coordinate, and the exponential term dependent on the relative coordinate, in complete analogy to Eq.~(\ref{eq:para}). The explicit calculation of the transition amplitudes regarding the internal motion can be found in Ref.~\cite{landau}. The main advantage of this treatment is that the effects of the electron spin are taken into account both directly and properly.  Such a relativistic treatment does not change the scalar factors  which couple atomic internal transitions to the electromagnetic field; they are proportional to $Y_{\ell,-m_\ell}$. An important modification is found in the polarization factor that in the non relativistic case is encoded in $\mathbf{e}_{m_{ph}}$, and that now,  distinguishes electric and magnetic effects through the presence of either electric or magnetic spherical polarization vectors \cite{landau}.

\section{Center of mass transition integrals $ I_{CM}$.}

Since both the internal and center of mass states are simultaneously coupled to the electromagnetic field,
controlling transitions between internal atomic states requires the proper regulation of center of mass transitions.
During the last years, there has been impressive developments on laser cooling and trapping techniques, as well as
an acquired capability to implement electric and magnetic fields on design. This has lead to the possibility of
creating atomic samples in which each atom can be properly described by a wave function of precisely the form
\begin{equation}\Psi(\mathbf{R},\mathbf{r}_{rel}) = \sum_{i,j}c_{ij}\Phi_{\kappa^{CM}_i}({\bf R})\phi_{\kappa^{rel}_j}({\bf r}_{rel}).\label{eq:super}\end{equation}
In particular, this applies to some experiments of few ions in Paul and Penning traps\cite{Wine} and, very recently, to neutral atoms magnetically trapped in electromagnetic cavities \cite{rempe}. In the case of ultra cold atomic gases in the degenerate regime, and for the condensate fraction, Eq.~(\ref{eq:super}) can be used with $\Phi_{\kappa^{CM}_i}$ determined, e. g., by the Gross-Pitaevski equation for boson samples.

In most other cases, environmental effects cannot be ignored and a description in terms of a density matrix for the center of mass state would be the most realistic option. The main reason behind this fact is the small value of the energy difference between the center of mass states of an isolated atom, compared to the interaction energy associated to collisions with other atoms and to the interaction with the background electromagnetic field. The implementation of semiclassical approaches where the center of mass motion is described in terms of Newton equations, while the internal dipole transitions are dealt within a quantum approach has probed to be
adequate for the description of recoil effects in most standard situations \cite{lembessis,letokhov,cohen2,ashkin}.  An alternative could be to take, as starting point, an idealized configuration in which each atom behaves as in empty free space with the center of mass stationary states  taken as  plane wave (PW) functions. Then the center of mass integral
$ I_{CM}$, Eq.~(\ref{eq:ICM}), can be evaluated
\begin{equation}
I^{PW}_{CM}(\mathbf{k};\mathbf{k}^{CM}_0;\mathbf{k}^{CM}_F) = \delta (\mathbf{k} - \Delta\mathbf{k}_{CM}),\quad \quad
\Delta\mathbf{k}_{CM} = \mathbf{k}^{CM}_F - \mathbf{k}^{CM}_0 \label{eq:cmpw}.
\end{equation}
The corresponding transition amplitudes, Eq.~(\ref{eq:TOF}) (which is expected to be dominant), Eq.~(\ref{eq:HI2}) and Eq.~(\ref{eq:HI3}) would  incorporate elementary recoil effects. The resulting transition rates should then be  introduced in Bloch equations where other properties like collision and temperature effects could also be taken into account. Such a calculation is beyond the scope of this work. Nevertheless, in the following section, the consequences of using
the center of mass factors $I^{PW}_{CM}$  in the evaluation of the transition amplitudes is discussed for structured modes.

Since ultra cold atoms are confined into a given region of space, in many cases, a reasonable description of
the center of mass states is given by  harmonic oscillator states centered at a position $\mathbf{R_0}$. Then,
the center of mass wave function has an structure $\Phi^*_{\kappa^{CM}_0}({\bf R}-\mathbf{R_0})$ and the
transition integral $ I_{CM}$, Eq.~(\ref{eq:ICM}), takes the form
\begin{eqnarray}
I_{CM}(\mathbf{k};\kappa^{CM}_0;\kappa^{CM}_F)&=& \int d^3R\Phi^*_{\kappa^{CM}_0}({\bf R}-\mathbf{R_0}) e^{i\mathbf{k}\cdot R}\Phi_{\kappa^{CM}_F}({\bf R}-\mathbf{R_0})\nonumber\\
&=& e^{i\mathbf{k}\cdot \mathbf{R}_0}\int d^3R\Phi^*_{\kappa^{CM}_0}({\bf R}) e^{i\mathbf{k}\cdot R}\Phi_{\kappa^{CM}_F}({\bf R})\label{eq:icm}
\end{eqnarray}
The exponential factor $e^{i\mathbf{k}\cdot \mathbf{R}_0}$ can also be written in terms of the wave vector $\mathbf{k}$ and the position of the center of the trap $\mathbf{R}_0$
using again Eq.~(\ref{eq:sph}). Notice however that for the center of mass coordinate, $\omega R_0/c$ may be greater than one.
For the particular case of an atom confined in a potential that has either spherical symmetry or cylindrical symmetry around the $z$-axis the integral in Eq.(\ref{eq:icm}) can be directly evaluated (see Appendix C). In the first case the dependence on $\hat{\mathbf{k}}$ is through the spherical harmonics:
 \begin{equation}
I^{ST}_{CM}(\mathbf{k};\kappa^{CM}_0;\kappa^{CM}_F) = {\mathcal{I}}^{ST}_{CM}(\hbar\omega^2/M_T\omega_{CM}c^2;\kappa^{CM}_0;\kappa^{CM}_F;\ell)Y_{\ell_{CM},m_{CM}}(\hat{\mathbf{k}}). \label{eq:cmYlma}
\end{equation}

Finally, if the center of was motion is described in terms of free spherical waves:
 \begin{equation}
I^{SW}_{CM}(\mathbf{k};\kappa^{CM}_0;\kappa^{CM}_F) = {\mathcal{I}}^{SW}_{CM}(\omega\lambda_{dB}/c;\kappa^{CM}_0;\kappa^{CM}_F;\ell)Y_{\ell_{CM},m_{CM}}(\hat{\mathbf{k}}). \label{eq:cmYlmb}
\end{equation}
with $\lambda_{dB}$ the de Broglie wavelength of the center of mass.
 The vector integrals $\mathbf{I}_{CM}=\nabla_{\mathbf{k}}I_{CM}$, Eq.~(\ref{eq:ICM2}), in both cases
can be directly written in terms of vectorial spherical harmonics (Appendix B).

\section{Transition amplitudes $\mathrm{T}_{0F}$ for a center of mass description in terms of plane waves.}
If the center of mass factor $I_{CM}$ is given by Eq.~(\ref{eq:cmpw}), then
\begin{equation}
\mathrm{T^{PW}_{0F}}
=\int d\omega S(\omega) e^{-i\Delta t}
  \delta(\Delta k^{CM}_z -\sqrt{\frac{\omega^2}{c^2} - (\Delta k_\bot^{CM})^2})
 \mathbf{J}^+_{rel}(\Delta\mathbf{k}^{CM};\kappa^{rel}_0;\kappa^{rel}_F)\cdot {\mathbf{e}^{(b)}_\kappa} \tilde{\Psi}_{\kappa;\omega}^{(b)}(\Delta k^{CM}_x,\Delta k^{CM}_y)
\end{equation}
For electric dipole (ED) transitions,
\begin{eqnarray}
\mathrm{T^{PW:ED}_{0F}}
= \frac{1}{\sqrt{4\pi}}\sum_{m_{ph}}&&\int d\omega S(\omega) e^{-i\Delta t}
  \delta(\Delta k^{CM}_z -\sqrt{\frac{\omega^2}{c^2} - (\Delta k_\bot^{CM})^2})\nonumber \\
&\cdot&\mathfrak{R}^{(1)}_{0,0,\kappa,\kappa^\prime}(\mu\omega/c M_e)
\upsilon^{(4)}(l,l^\prime,0,1;m,m^\prime,0,m_{ph})\hat{\mathbf{e}}^*_{m_{ph}}\cdot {\mathbf{e}^{(b)}_\kappa} \tilde{\Psi}_{\kappa;\omega}^{(b)}(\Delta k^{CM}_x,\Delta k^{CM}_y),
\end{eqnarray}
which is directly proportional to the angular spectrum as could be expected. This equation is a quantum analog in momentum space to the standard dipole interaction term $-\mathbf{d}\cdot\mathbf{E}(\mathbf{R})$ used in most
semiclassical descriptions of ED recoil effects. From this equation we observe that recoil effects are
determined directly by the polarization and angular spectrum of the mode under consideration.
This result has direct physical consequences, $e.$ $g.$, a Bessel mode \cite{jauregui} or a Laguerre-Gauss mode \cite{Davila,alexandrescu} can induce center of mass rotations around the main propagation axis of the beam, while a Weber mode has probed to be adequate for inducing a center of mass motion with parabolic symmetry on cold atomic samples \cite{joszef}.

For electric quadrupole (EQ) and magnetic dipole (MD) transitions the transition amplitude has the structure
\begin{eqnarray}
\mathrm{T^{PW:A}_{0F}}
&=&\sum_{m_{ph},m_1}\int  d\omega S(\omega) e^{-i\Delta t}
  \delta(\Delta k^{CM}_z -\sqrt{\omega^2/c^2 - (\Delta k_\bot^{CM})^2})\nonumber\\
&\cdot&\mathfrak{R}^{(1)}_{1,m_1,\kappa,\kappa^\prime}(\mu\omega/c M_e)
\upsilon^{(4)}_A(l,l^\prime,1,1;m,m^\prime,m_1,m_{ph})\hat{\mathbf{e}}^*_{m_{ph}}\cdot {\mathbf{e}^{(b)}_\kappa} Y_{1,m_1}(\hat{\Delta\mathbf{k}}^{CM})\tilde{\Psi}_{\kappa;\omega}^{(b)}(\Delta k^{CM}_x,\Delta k^{CM}_y),
\end{eqnarray}
where the EQ or MD character of the transition is encoded in the factor $\upsilon^{(4)}_A$ which  depends on
both the mode polarization coupling parameter $m_{ph}$ and the effective angular momentum factor $m_1$. In this case, the effective interaction term yielding recoil effects depends on the polarization and products of the angular spectrum of the mode with first order terms of $\mathbf{k}$. In configuration space, these products correspond to first order derivatives of the electric and magnetic fields.

Due to its frequent use in actual experiments, we discuss in more detail axial symmetric scalar Gaussian modes of waist $\mathrm{w}_\bot$,
$$\tilde \Psi_{\mathrm{w}_\bot}(\mathbf{k}_\bot) = {\mathcal{N}_{\mathrm{w}_\bot}} \mathrm{e}^{-\mathrm{w}_\bot^2 k_\bot^2}.$$
For scalar circularly polarized modes, the exact expression of the  normalization integral is evaluated in Appendix A,
\begin{equation}
\vert{\mathcal{N}_{\mathrm{w}_\bot}}\vert^2 = \hbar\omega\Big[\frac{\omega^2\pi^3}{c^2}\big[\frac{1}{2a^2} +\sqrt{2\pi}\frac{4a^2 -1}{8 a^3}e^{-2a^2} \mathrm{Erfi}(\sqrt{2}a)\big]\Big]^{-1},\quad \quad a =\frac{\omega \mathrm{w}_\bot}{c},\label{eq:ngauss}
\end{equation}
with Erfi the imaginary error function $\mathrm{Erfi}(z)= \mathrm{erf}(iz)/i.$ The factor $\hbar\omega$ is a natural scale of
the energy of the mode that could be used for the quantization of the EM field. For our study, it just provides a reference useful for comparing the effects of different structured modes with  the same  EM energy on atomic transitions rates.
For Gaussian modes, different choices of the waist   $\mathrm{w}_\bot$ modify the relevance of the  $Y_{1,m_1}(\hat{\Delta\mathbf{k}}^{CM})$ factor, Eq.(\ref{eq:Y1m}); that is, for paraxial modes $$Y_{1,0}(\hat{\Delta\mathbf{k}}^{CM})\sim\frac{1}{\sqrt{4\pi}},\quad\quad Y_{1,\pm 1}(\hat{\Delta\mathbf{k}}^{CM})\sim 0$$
so that the internal transition rates which would be dominant are those with  $\upsilon^{(4)}_A(l,l^\prime,1,1;m,m^\prime,0,m_{ph})\ne 0$. In the antiparaxial regime $k_\bot\sim \omega/c$, that is $\mathrm{w}_\bot\rightarrow 0$,
$$Y_{1,0}(\hat{\Delta\mathbf{k}}^{CM})\sim 0,\quad\quad Y_{1,\pm 1}(\hat{\Delta\mathbf{k}}^{CM})\sim \frac{e^{\pm i\varphi_{\mathbf{k}_{CM}}}}{\sqrt{8\pi}}$$
and  the internal transition rates which would be dominant are those with  $\upsilon^{(4)}_A(l,l^\prime,1,1;m,m^\prime,\pm 1,m_{ph})\ne 0$. Notice however, that for a Gaussian mode the transverse momenta distribution is always centered around $\mathbf{k}_\bot = \mathbf{0}$
discarding them as the optimal choice for achieving the antiparaxial regime. A better option corresponds to using modes with a $\mathbf{k}_\bot$ distribution that becomes maximum at a $k_\bot$ value different from zero, like the propagation invariant modes
mentioned at Section II.

\section{Transition amplitudes $\mathrm{T}_{0F}$ for atoms trapped in a spherically symmetric potential.}
If the atom is trapped in a spherically symmetric external potential of natural frequency $\omega_{CM}$, the structure of the scalar center of mass term $I_{CM}$ in the transition amplitude is given by Eq.~(\ref{eq:cmYlma}), while that of vectorial internal transition factor $\mathbf{J}^+_{rel}$ has the structure given by Eq.~(\ref{eq:Jplus}). Since the product of two spherical harmonics can always be written as a superposition of spherical harmonics, all the transition amplitudes $\mathrm{T_{0F}}$ can then be  written in terms of the basic integrals
\begin{equation}
\mathcal{H}^{(b;m_{ph};\kappa)}_{\ell,m_\ell} = \int d^2{\mathbf{k}}_\bot \hat{\mathbf{e}}^*_{m_{ph}}\cdot {\mathbf{e}^{(b)}_\kappa} \tilde \Psi_{\kappa}^{(b)}(\mathbf{k}_\bot)Y_{\ell,m_\ell}(\hat{\mathbf{k}}).\label{eq:hbmk}
\end{equation}

 Similar arguments apply to the vectorial factor $\mathbf{I}_{CM}$ and the scalar factors $J^{\pm}_{rel}$, so that these basic integrals also determine the transition amplitudes associated to the interaction Hamiltonians $\hat H_{I2}$, Eq.~(\ref{eq:HI2}) and  $\hat H_{I3}$, Eq.~(\ref{eq:HI3}).

In general, any integral $\mathcal{H}^{(b;m_{ph};\kappa)}_{\ell,m_\ell}$
is relevant for the evaluation of more than one transition amplitude.
For instance, internal electric dipole transitions induced by $\hat H_{I1}$ involve $Y_{00}(\hat{\mathbf{k}})$ which may be coupled
to a center of mass transition  with no  rotational effects, that is $\ell_0^{CM} =\ell_F^{CM}$
and $m_0^{CM} =m_F^{CM}$. In such a case, the center of mass term involves a term $Y_{00}(\hat{\mathbf{k}})$,
and the integral $\mathcal{H}_{0,0}$ should be evaluated. A very different situation corresponds to an
electric quadrupole transition induced also by $\hat H_{I1}$ which involves  internal $Y_{1,m}(\hat{\mathbf{k}})$ terms and a center of mass transition
with a similar contribution (this transition could be accompanied by a change in the rotational wave numbers of the center of mass motion). The coupling of the two spherical harmonics will again have a contribution of the integral $\mathcal{H}_{0,0}$, but with different internal and center of mass  factors.

 In the following subsections we elaborate on the dependence of the $\mathcal{H}^{(b;m_{ph};\kappa)}_{\ell,m_\ell}$ on the electromagnetic parameters $b$ and $\kappa$ for different EM modes.

\subsection{Spherical EM modes.}
Electromagnetic fields with spherical symmetry can be written in terms of vector spherical harmonics as described in Appendix B. If recoil effects are ignored in a radiative transition, the evaluation of internal transition atomic probabilities (including both electric and magnetic effects important in the  relativistic realm) can be naturally described in terms of these modes  \cite{landau}. In this section, contrary to standard calculations,  recoil effects have been incorporated in the calculation.
 The
integrals $\mathcal{H}^{(b;m_{ph};\kappa)}_{\ell,m_\ell}$ necessary to study the  role for a given  EM spherical mode ${\mathbb{Y}}_{jm}^{(b)}$  on the coupled angular momentum changes in the  internal and center of mass motion (as encoded in a $Y_{\ell m_\ell}\hat{\mathbf{e}}^*_{m_{ph}}$ with $m_{ph}$ = 0, $\pm$ 1) are

\begin{equation}
{\mathcal{H}}^{(b;m_{ph};jm)}_{\ell,m_\ell} = \frac{\sqrt{\hbar\omega}}{2\pi} \int d^2\mathbf{k}_\bot \hat{\mathbf{e}}^*_{m_{ph}}\cdot
{\mathbb{Y}}_{jm}^{(b)}(\hat{\mathbf{k}}) Y_{\ell m_\ell}(\hat{\mathbf{k}})
=\frac{\sqrt{\hbar\omega^3}}{2\pi c}
\int d\Omega_{\hat{\mathbf{k}}}
\Big({\mathbb{Y}}_{jm}^{(b)}(\hat{\mathbf{k}})\Big)_{m_{ph}}
Y_{\ell m_\ell}(\hat{\mathbf{k}}).\label{eq:sphH}
\end{equation}
 Note that, the vector integrals $\mathbf{I}_{CM}=\nabla_{\mathbf{k}}I_{CM}$, Eq.~(\ref{eq:ICM2}), are orthogonal to magnetic vector modes,  the terms of the Hamiltonian $\hat H_{I2}$ that involve $\mathbf{I}_{CM}$ would yield center of mass transitions just in the case that electric spherical modes are used to induce them.

Using  equations that can be found in  Appendix B, it can be shown that
\begin{eqnarray}
\vert {\mathcal{H}}^{(E;{m_{ph}};jm)}_{\ell,m_\ell}\vert &=&\frac{\sqrt{\hbar\omega^3}}{2\pi c}
\Big[ \delta_{\ell,j+1}\delta_{m_\ell,m+{m_{ph}}}\sqrt{j} \begin{pmatrix}j+1 & 1 & j\cr m+{m_{ph}} &-{m_{ph}} &-m\end{pmatrix}\nonumber\\
&+& \delta_{\ell,j-1}\delta_{m_\ell,m+{m_{ph}}}\sqrt{j+1}\begin{pmatrix}j-1 & 1 & j\cr m+{m_{ph}} &-{m_{ph}} &-m\end{pmatrix} \Big]\nonumber\\
\vert {\mathcal{H}}^{(M;{m_{ph}};jm)}_{\ell,m_\ell}\vert &=&\frac{\sqrt{\hbar\omega^3}}{2\pi c}
\delta_{\ell,j}\delta_{m_\ell,m+{m_{ph}}}\sqrt{2j+1}\begin{pmatrix}j & 1 & j\cr m+{m_{ph}} &-{m_{ph}} &-m\end{pmatrix},
\end{eqnarray}
here, $E$ and $M$ denote electric and magnetic spherical vector waves. The $\delta$ factors indicate which
spherical waves induce the coupled (internal and center of mass) atomic transitions.
Taking into account the behavior of the 3j-symbols, in general,
\begin{equation}\sqrt{\frac{c^2}{\hbar\omega^3}}\vert {\mathcal{H}}^{(b;{m_{ph}};jm)}_{\ell,m_\ell}\vert \le \frac{1}{2\pi},\quad\quad b = E,M. \label{eq:sphh}\end{equation}
This result will be useful for the clear comparison between the integrals $\mathcal{H}^{(b;m_{ph};\kappa)}_{\ell,m_\ell}$ for different selections of the  electromagnetic modes.

Though spherical EM modes give a natural scenario for the description of spontaneous internal transition rates,
induced atomic transitions using these modes would require an efficient procedure for their selected generation; this is a non trivial task. Perhaps the simplest implementation could be achieved by the controlled excitation of predetermined EM modes within cavities with spherical symmetry.

\subsection{Gaussian modes.}
 For axial symmetric scalar Gaussian modes of waist $\mathrm{w}_\bot$, the integrals of the EM field  that go along the
polarization vectors $\hat{\mathbf{e}}_\pm$ necessary to evaluate atomic transition amplitudes are
\begin{equation}
\mathcal{H}^{(\pm,\kappa)}_{\ell,m_\ell} =\delta_{m,0} {\mathcal{N}_{\mathrm{w}_\bot}}\sqrt{(2\ell +1)\pi}\frac{\omega^2}{c^2}
\int_0^1 dt  P_\ell(\sqrt{1-t^2}) e^{-a^2 t^2}t,
\end{equation}
while the integrals of the EM field  that go along the
polarization vector $\hat{\mathbf{e}}_z$ are
\begin{equation}
\mathcal{H}^{(z,\kappa)}_{\ell,m_\ell} =\delta_{m,0} {\mathcal{N}_{\mathrm{w}_\bot}}\sqrt{(2\ell +1)\pi}\frac{\omega^2}{c^2}
\int_0^1 dt  P_\ell(\sqrt{1-t^2}) e^{-a^2 t^2}\sqrt{1-t^2}.
\end{equation}
Some specific expressions are:
\begin{eqnarray}
\mathcal{H}^{(\pm,\mathrm{w}_\bot)}_{0,0} &=&{\mathcal{N}_{\mathrm{w}_\bot}}\sqrt{\pi}\frac{\omega^2}{c^2}\frac{1 - e^{-a^2}}{2 a^2} \nonumber \\
\mathcal{H}^{(\pm,\mathrm{w}_\bot)}_{1,0} &=&{\mathcal{N}_{\mathrm{w}_\bot}}\sqrt{3\pi}\frac{\omega^2}{c^2}\frac{2a -\sqrt{\pi} e^{-a^2}\mathrm{Erfi}(a)}{4 a^3} \nonumber \\
\mathcal{H}^{(\pm,\mathrm{w}_\bot)}_{2,0} &=&{\mathcal{N}_{\mathrm{w}_\bot}}\sqrt{5\pi}\frac{\omega^2}{c^2}\frac{2a^2 -3 + \mathrm{e}^{-a^2}(  3+a^2)}{4a^4}.
\end{eqnarray}
These functions are plotted in Figure 1. It can be observed that it is
necessary to consider modes with values of $a=\omega \mathrm{w}_\bot/c\sim 1$ to observe a substantial increment of the transition rates with respect to
the result obtained for paraxial beams. In general, the maxima  of $\vert H^{(\pm,\mathrm{w}_\bot)}_{\ell,m}\vert$ are achieved for waists $\mathrm{w}_\bot$ smaller than the wavelength. As it could be expected, the functions $H^{(z,\mathrm{w}_\bot)}_{\ell,m}$ decay as $\mathrm{w}_\bot$ increases faster than the functions  $H^{(\pm,\mathrm{w}_\bot)}_{\ell,m}$ do.
In fact,for $a>>1$
\begin{equation}
\mathcal{H}^{(\pm,\mathrm{w}_\bot)}_{\ell,0}\rightarrow \frac{\sqrt{2(2\ell +1)}}{\pi}\frac{(2m)!}{2^\ell m!m^\prime!}\sqrt{\hbar\omega} \frac{1}{\mathrm{w}_\bot},\label{asympt:gauss}
\end{equation}
$m=m^\prime=\ell/2$ if $\ell$ is even, and $m = (\ell -1)/2 = m^\prime -1$ if $\ell$ is odd.
As a consequence for a scalar Gaussian mode, taking into account the normalization factor given in Eq.(\ref{eq:ngauss}), all transition rates depend on the energy per unit transverse area of the EM mode in the paraxial limit.

We have also evaluated the integrals $\mathcal{H}^{(b;m_{ph};\mathrm{w}_\bot)}_{\ell,m}$ for transverse electric and transverse magnetic Gaussian modes. The integrands differ from those of scalar beams by  factors $k_\bot c/\omega$, $k_\bot k_zc^2/\omega^2$
and $k_\bot^2c^2/\omega^2$ as a consequence of Eq.~(\ref{eq:polarizations}) and Eq.~(\ref{eq:hbmk}). They also decay as $\mathrm{w}_\bot^{-1}$ as $a>>1$, but they yield maxima with  higher values of $\mathcal{H}^{(\kappa,\mathrm{w}_\bot)}_{\ell,m}$  than the scalar modes. In fact some of these maxima are beyond the spherical modes limit, Eq.~(\ref{eq:sphh}).

\begin{figure*}
\begin{center}
\subfloat[]{\label{F:hgausss}\includegraphics[width=0.5\textwidth]{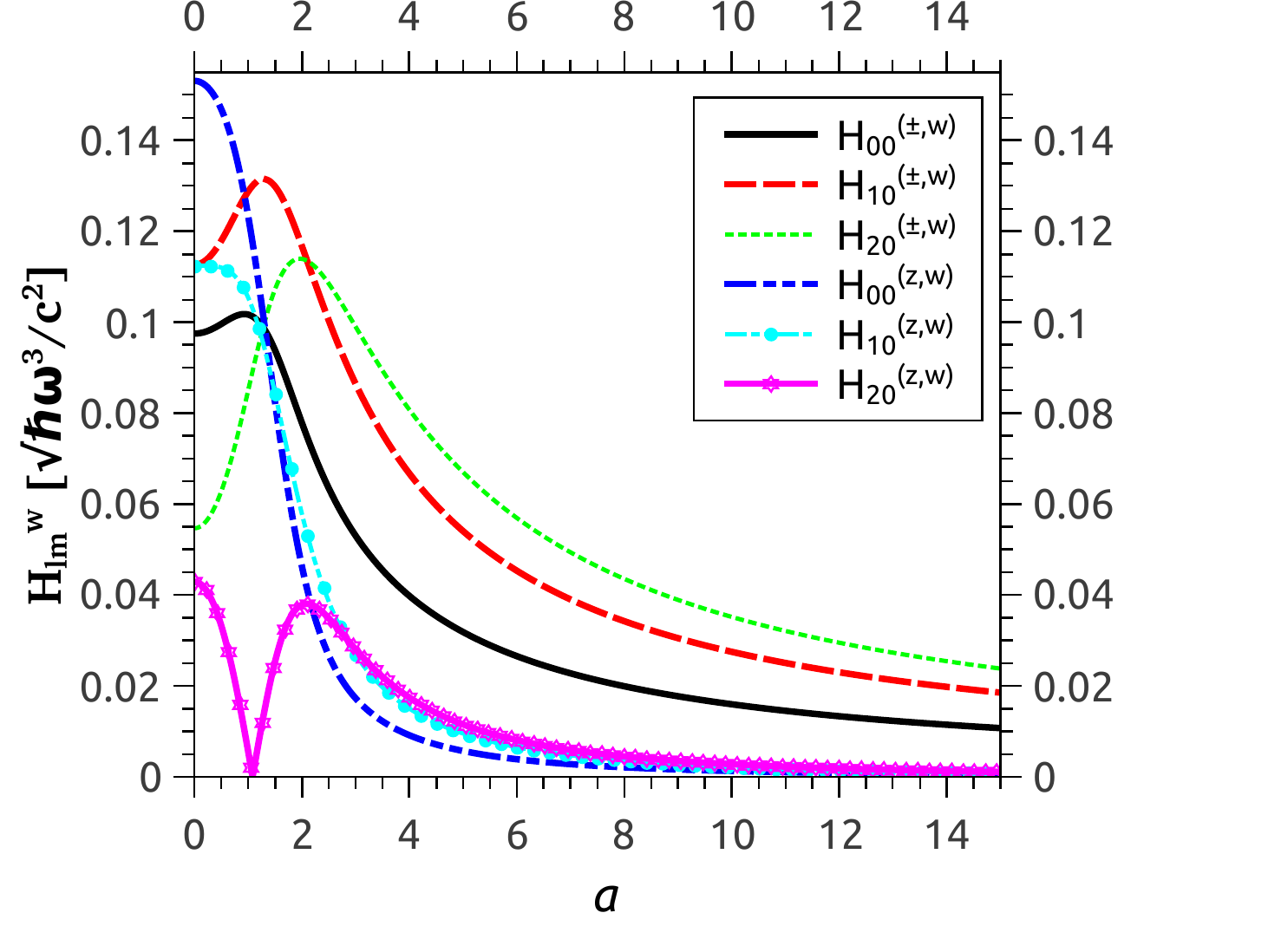}}
\subfloat[]{\label{F:hgaussv}\includegraphics[width=0.5\textwidth]{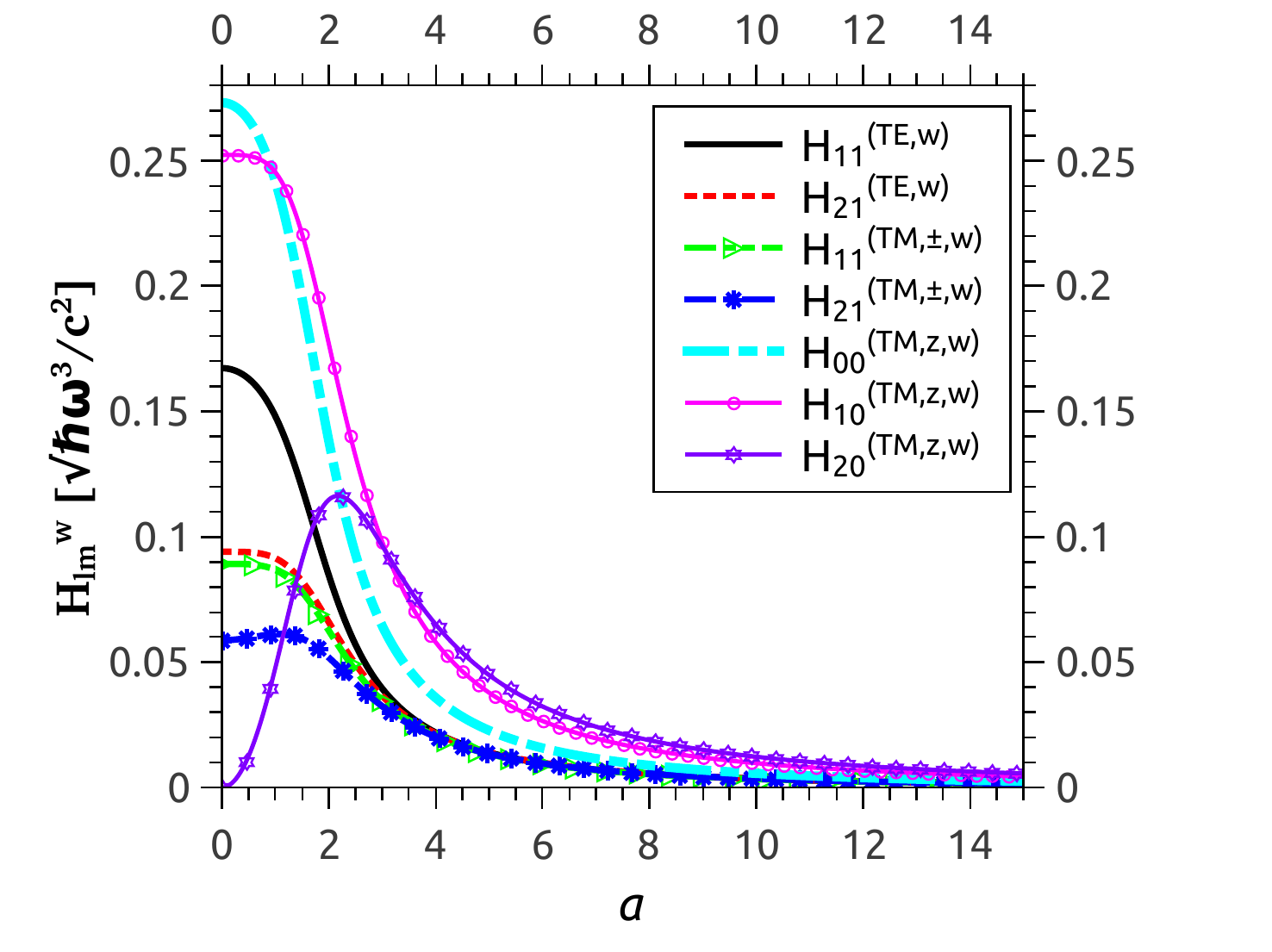}}
\end{center}
\caption{(Color  online)(a)Illustrative basic integrals $\mathcal{H}^{\mathrm{w}_\bot}_{l,m}$ for scalar Gauss modes; and (b) illustrative basic integrals $\mathcal{H}^{\mathrm{w}_\bot}_{l,m}$ for  vector TE and TM  Gauss modes. They are plotted as a function of their waist $\mathrm{w}_\bot$ measured in terms of the inverse wave number $a =
\omega \mathrm{w}_\bot/c$.  }
\end{figure*}

\subsection{Propagation invariant modes.}

In Appendix A, the normalization of propagation invariant modes is worked out in several schemes. Two of them refer to the case where ideal propagation modes are considered. The third is the closest to the experimental realizations, it corresponds to
work with a Gaussian superposition of ideal modes around a  given $\kappa_\bot$:
\begin{equation}
\tilde \Psi_{\kappa_\bot,\kappa_m:\mathrm{w}_\bot}^{(b)}(\mathbf{k}_\bot) = {\mathcal{N}^{(b)}_{\kappa_\bot,\mathrm{w}_\bot}} \mathrm{e}^{-\mathrm{w}_\bot^2 (k_\bot -\kappa_\bot)^2}\tilde\psi_{\kappa_m}
(\varphi_{\mathbf{k}}),\label{gauss:piof}
\end{equation}
with $\mathrm{w}_\bot^{-1}\gg\kappa_\bot$.
The relevant integrals for scalar propagation invariant modes can then be written as
\begin{eqnarray}
\mathcal{H}^{(b,\kappa_\bot)}_{\ell,m_\ell} &=& h^{(b,\kappa_\bot)}_{\ell,m_\ell}\Theta^{(\kappa_m)}_{m_\ell}\nonumber\\
\tilde h^{(b,\kappa_\bot)}_{\ell,m_\ell} &=& \sqrt{\frac{(2\ell +1)}{2}\frac{(\ell-\vert m_\ell\vert)!}{(\ell+\vert m_\ell\vert)!}}\frac{\omega^2}{c^2}
\int_0^1 dt {\mathcal{N}^{(b)}_{\kappa_\bot,\mathrm{w}_\bot}} P^{\vert m_\ell\vert}_\ell(\sqrt{1-t^2}) e^{-a^2 (t-{\tilde{t}})^2}t, \quad\quad \tilde{t} = \kappa_\bot c/\omega\nonumber\\
\tilde \Theta^{(\kappa_m)}_{m_\ell}&=& \frac{1}{\sqrt{2\pi}}\int_0^{2\pi}d\varphi_{\mathbf{k}} \tilde\psi_{\kappa_m}(\varphi_{\mathbf{k}})\mathrm{e}^{im_\ell\varphi}
\end{eqnarray}
The set of terms $h^{\mathrm{w}_\bot}_{l,m}$ is common to all propagation invariant modes, while $\Theta^{(\kappa_m)}_{m_\ell}$
depends on the details of the  angular spectra that characterizes each kind of propagation invariant modes.

\begin{figure*}
\begin{center}
\subfloat[]{\label{F:piofscalar}\includegraphics[width=0.5\textwidth]{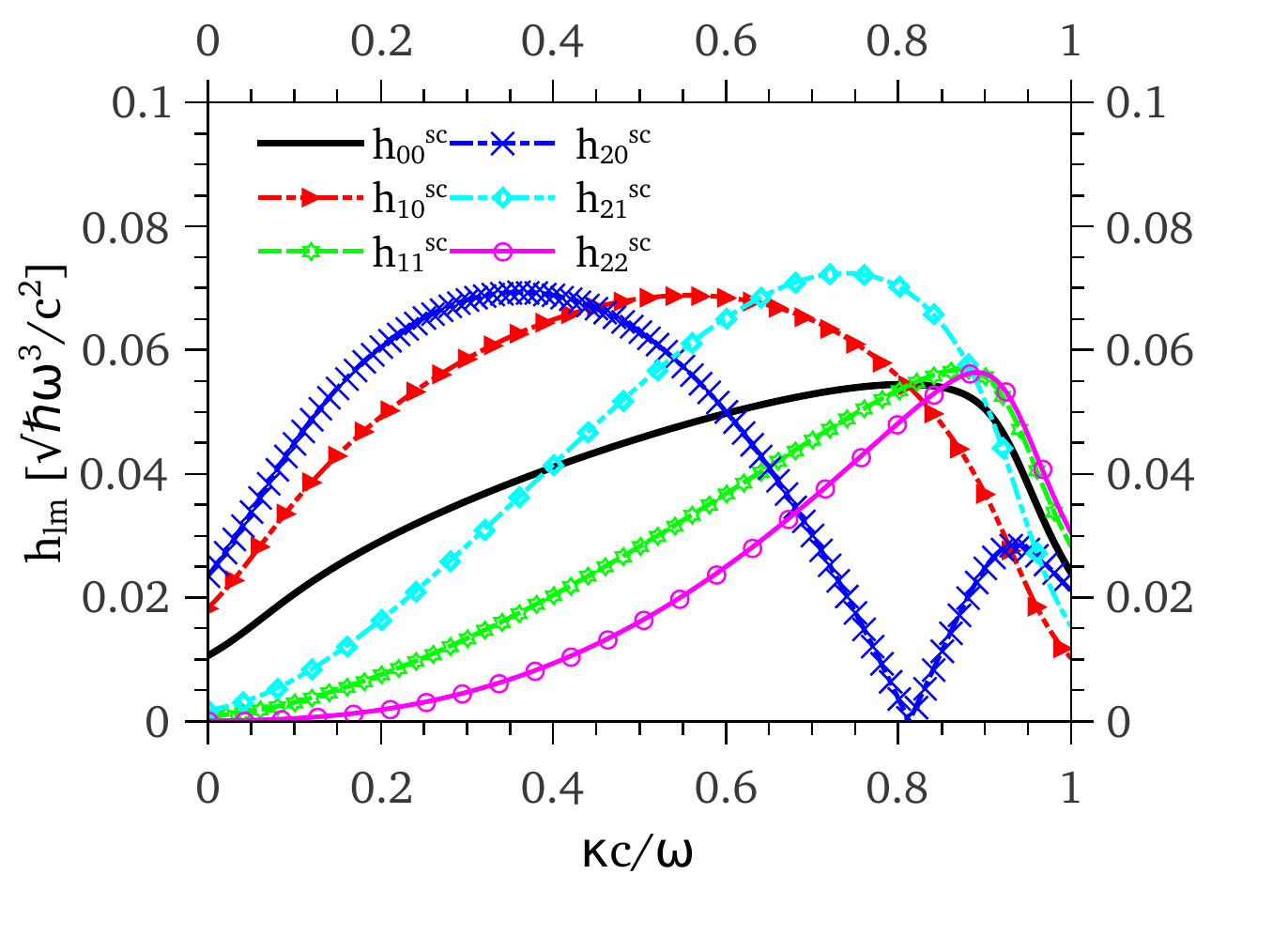}}
\subfloat[]{\label{F:piofvector}\includegraphics[width=0.5\textwidth]{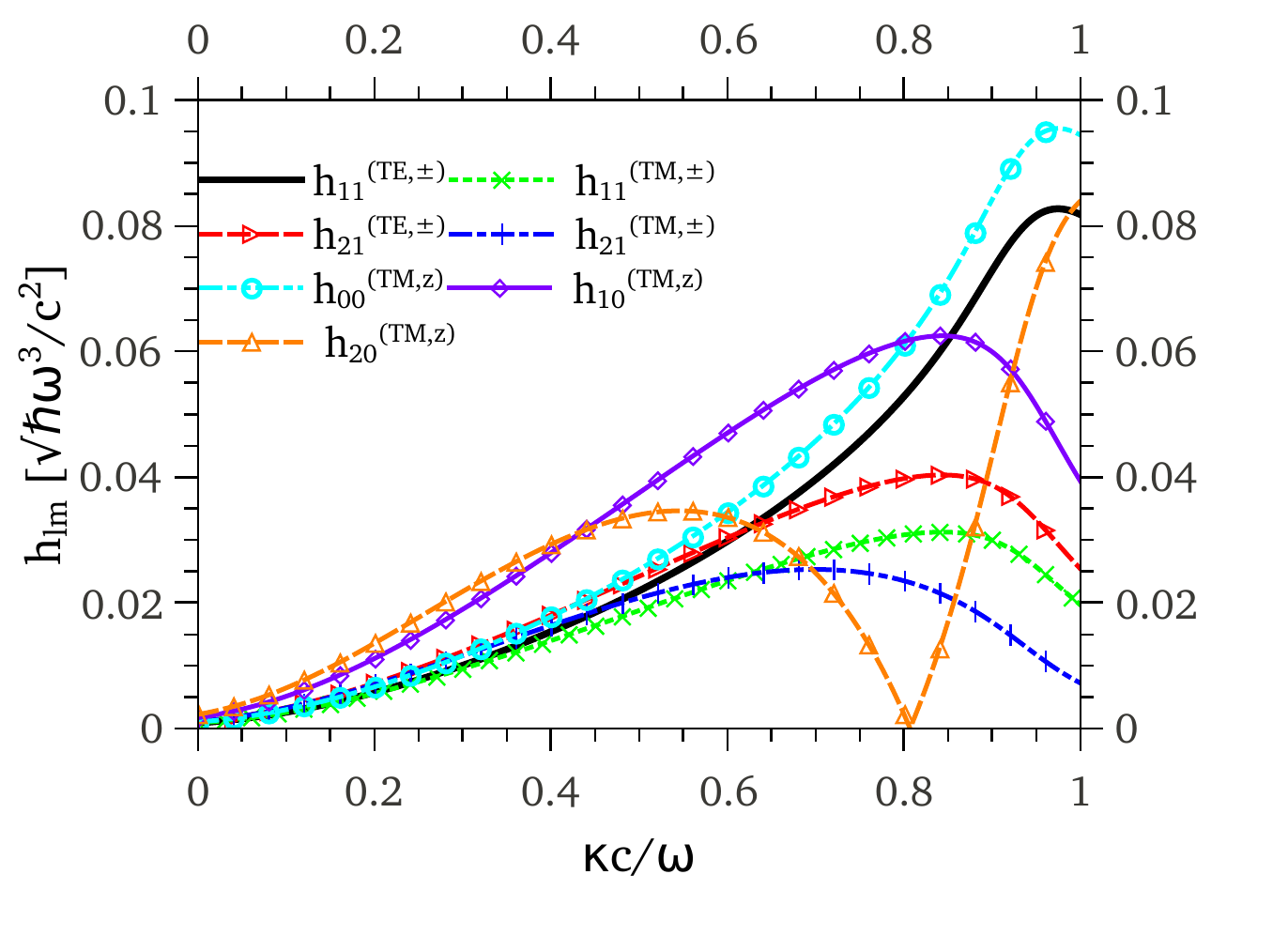}}
\end{center}
\caption{(Color  online) Basic $k_\bot$ integrals, $h^{\mathrm{w}_\bot}_{l,m}$ for (a) scalar propagation invariant modes (b) transverse electric and transverse magnetic propagation invariant modes as a function of the transverse wave number $\kappa_\bot$ measured in units of $\omega/c$. The waist of the Gaussian factor was taken as $10 c/\omega.$ }
\end{figure*}

Figure \ref{F:piofscalar} illustrates the absolute value of the integrals $\vert\tilde h^{\mathrm{w}_\bot}_{\ell,m_\ell}\vert =  h^{\mathrm{w}_\bot}_{\ell,m_\ell} $ for scalar  modes as a function of the dominant transverse wave number $\kappa_\bot$ measured in units of $\omega/c$. In Fig~\ref{F:piofvector}, the vectorial analogues of these integrals are shown; their evaluation incorporate the terms $\hat{\mathbf{e}}^*_{ph}\cdot\mathbf{e}^{(b)}$ for $b= TE$, $TM$. Though both scalar and vector  integrals depend on the waist of the Gaussian $\mathrm{w}_\bot^{-1}$ used in Eq.~(\ref{gauss:piof}), the same reported general structure is observed whenever  $\mathrm{w}_\bot^{-1}\gg\kappa_\bot$. Note that there is a substantial increment of the $h^{\mathrm{w}_\bot}_{\ell,m_\ell}$ values as the parameter $\kappa_\bot$ deviates from the paraxial
approximation. These occurs both for scalar and vector beams.

The integrals $\tilde \Theta^{(\kappa_m)}_{m_\ell}$ have a very simple expression for Bessel beams:
\begin{equation}
\tilde \Theta^{(Bessel:m)}_{m_\ell} = \frac{1}{2\pi}\int_0^{2\pi}d\varphi_{\mathbf{k}} \mathrm{e}^{im\varphi_{\mathbf{k}}}\mathrm{e}^{im_\ell\varphi_{\mathbf{k}}}  = \delta_{-m,m_\ell}
\end{equation}
which is a manifestation of conservation of total angular momentum during the transition. Notice that this angular momentum can be originated from polarization terms like those in Eq.~(\ref{eq:polarizations}) or from orbital terms related to the angular spectra Eq.~(\ref{def:bessel:as}). They can also either be transferred to the center of mass or to the internal state of the atom depending
the origin of the $e^{i{m_\ell}\varphi_\bot}$ factor in the integral $\tilde \Theta^{(\kappa_m)}_{m_\ell}$.

For Mathieu beams $\tilde \Theta^{(Mathieu:m)}_{m_\ell}$ can be directly related to the $A_n^m$ and $B_n^m$ Mathieu coefficients in the standard notation\cite{nist}. They achieve a maximum value for $q=0$ for $m=m_\ell$ as could be expected since in this limit
the elliptic coordinates reduce to circular cylinder coordinates, and Mathieu modes Eq.~(\ref{def:mathieu:as}) become identical to
Bessel modes, Eq.~(\ref{def:bessel:as}). For moderate values of  $q$ and as it increases, the off diagonal terms $\tilde \Theta^{(Mathieu:m)}_{m_\ell}$, $m\ne m_\ell$ also increase. This means that Mathieu modes can be used to create superpositions of internal and center of mass states whenever the corresponding internal, ${J}_{rel}$ and center of mass, $I_{CM}$,  vector or scalar factors
have similar values for the ${m_\ell}$ values of interest.

\begin{figure*}
\begin{center}
{\label{F:mathieu}\includegraphics[width=0.5\textwidth]{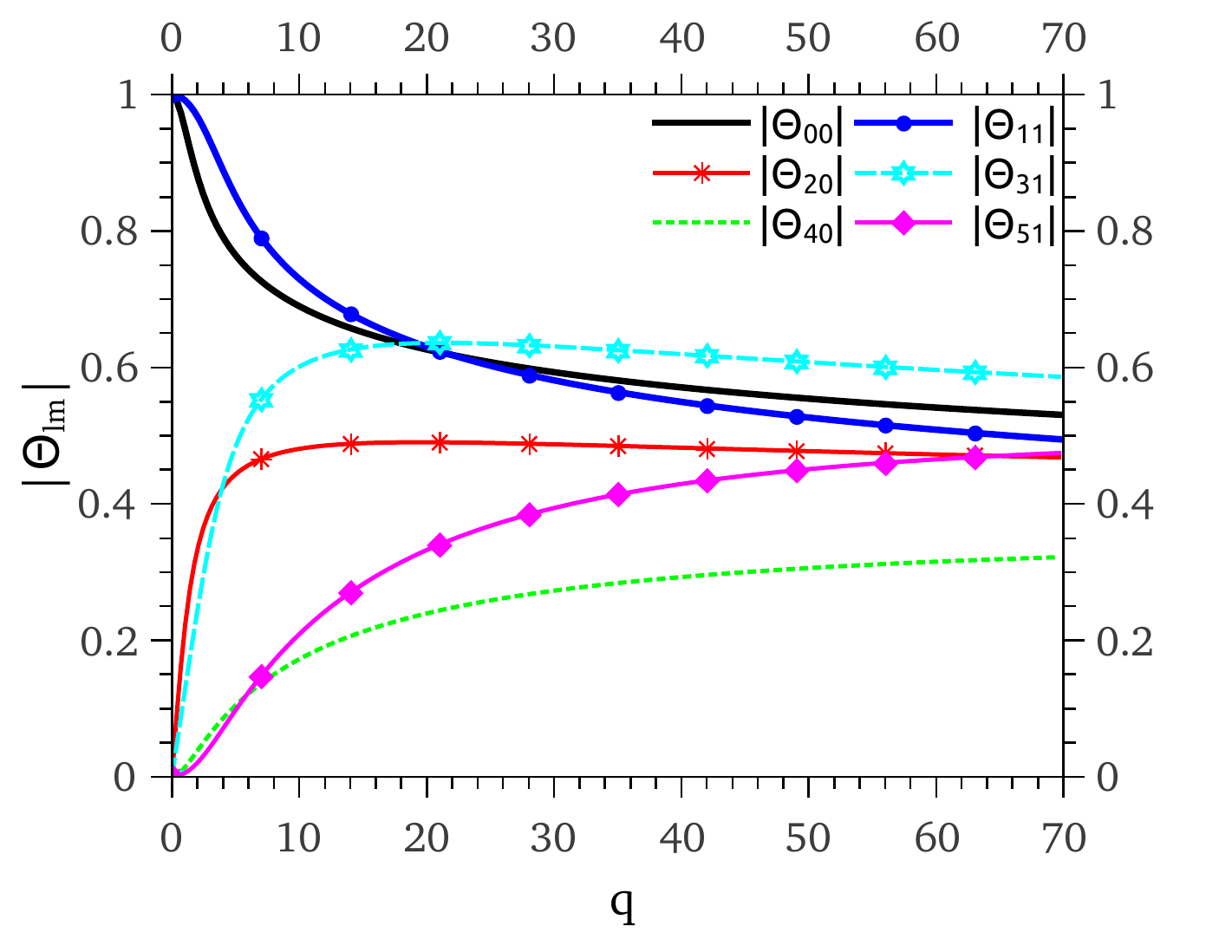}}
\end{center}
\caption{(Color  online) Absolute value of the angular spectrum factor $\vert \tilde \Theta^{(Mathieu:m)}_{m^\prime}\vert =\Theta_{mm^\prime}$ for Mathieu  modes as a function of the continuous Mathieu parameter $q =f\kappa_\bot/2 $ with $f$ the focal distance that defines the elliptic coordinates, and $\kappa_\bot$ the transverse wave number. }
\end{figure*}

We  now report the results for the absolute value of $\tilde \Theta^{(Weber:\alpha)}_{m_\ell}$ as a function of the continuous parameter $\alpha$ that defines the aperture of a Weber mode. In this case $\Theta^{(Weber:\alpha)}_{m_\ell}$ is much less than one, in general, with the exception of odd Weber beams with $\alpha$ nearby well identified discrete values.   As a consequence, if a transition involving  even Weber modes takes place, the resulting atomic states will be a superposition of many orbital angular momentum states. In general, due to the intrinsic central field basic structure of the internal atomic states, these means that the recoil effects of the transition on the atom will be better described not as simple rotational.  This is consistent with the fact that Weber modes are not eigenfunctions of the orbital angular momentum but carry, as a natural mechanical property, the product of the angular momenta along the $z$ axis and the linear momentum along a transverse direction  $p_y$, $L_zp_y$\cite{Miller}. For internal dipole transitions induced by $\hat H_{I1}$, the center of mass factor $I_{CM}$
 will have a maximum overlap with the angular spectrum of a Weber beam if
 $$I_{CM}\sim \frac{e^{i\alpha \ln\vert\tan\varphi_{\mathbf{k}}/2\vert}}{2\sqrt{\pi\vert\sin\varphi_{\mathbf{k}}\vert}}.$$
This could happen if the initial and final wave functions of the center of mass are solutions of the Schr\"odinger equation of a free particle with boundary conditions with parabolic symmetry. One important question is whether or not the corresponding transition can be described in terms of a conservation law involving the atomic and the electromagnetic dynamical property $L_zp_y$. A first experimental study with that scope was reported in Ref.~\cite{joszef}.

\begin{figure*}
\begin{center}
\subfloat[]{\label{we:ev}\includegraphics[width=0.5\textwidth]{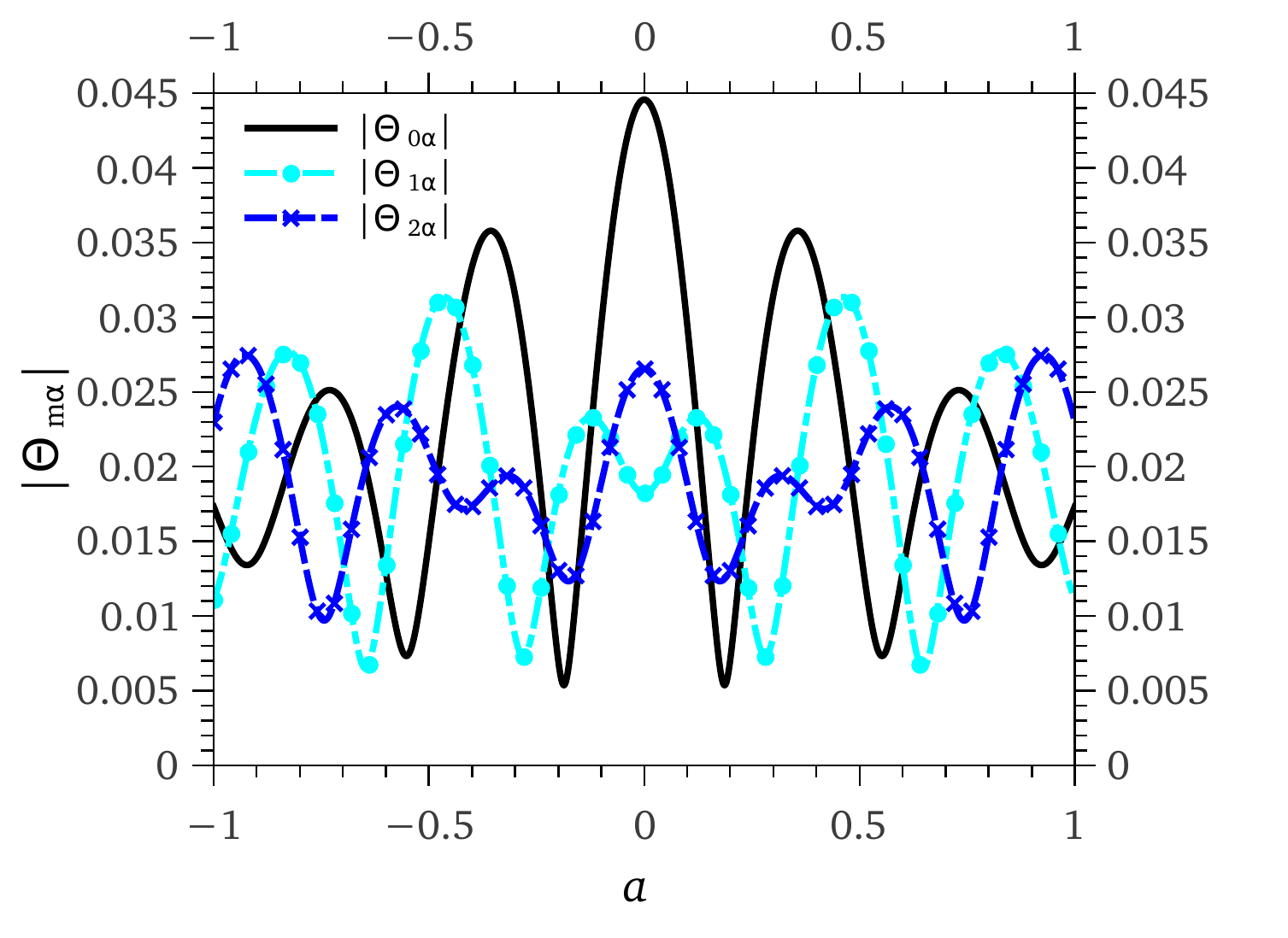}}
\subfloat[]{\label{we:odd}\includegraphics[width=0.5\textwidth]{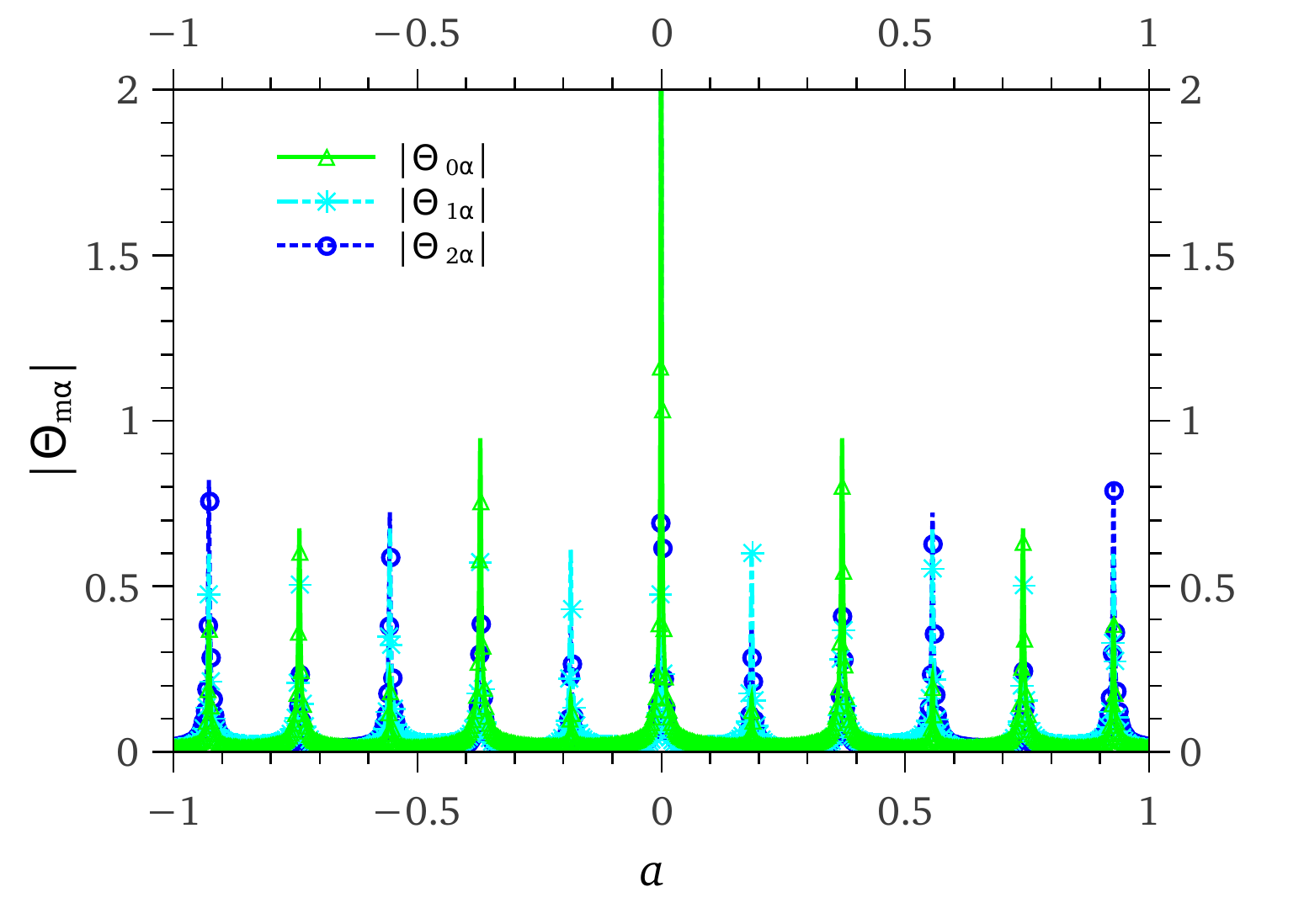}}
\end{center}
\caption{(Color  online) Absolute value of the angular spectrum factor $\vert \tilde{\Theta}^{(Weber:\alpha)}_{m_\ell}
\vert=\Theta^{(Weber:\alpha)}_{m_\ell}$ (a) for even Weber modes,  and (c) for odd Weber modes as a function of the continuous parameter $\alpha$. For all modes, $\Theta^{(Weber:\alpha)}_{m_\ell} = \Theta^{(Weber:\alpha)}_{-m_\ell}$.  }
\end{figure*}

\section{ Atoms in a trap with cylindrical symmetry interacting with either  Laguerre-Gaussian  or Bessel beams.}

For atoms in a cylindrical trap with the symmetry axis parallel to the main propagation axis of the light beam, according to the formalism developed in the first sections, the relevant integrals have the structure,
\begin{equation}
\mathcal{C}^{(b;\kappa)}_{\ell,m_\ell,m_{ph};\Delta m_R,\mu,\alpha_\rho,\alpha_z} = \int d^2{\mathbf{k}}_\bot{\mathcal{R}}^{\Delta m_{R},\mu}({k_\bot\alpha_\rho}) {\mathcal{Z}}^{n_z,n^\prime_z}(\alpha_z\sqrt{(\omega^2/c^2-k_\bot^2})\Big[\hat{\mathbf{e}}^*_{m_{ph}}\cdot {\mathbf{e}^{(b)}_\kappa}\Big] \tilde \Psi_{\kappa}^{(b)}(\mathbf{k}_\bot)Y_{\ell,m_\ell}(\hat{\mathbf{k}}).\label{eq:cbmk}
\end{equation}
For the $\mathrm{T}_{0F}$ matrix element,
 \begin{equation}
\mathrm{T}_{0F} =\eta_{rel}\frac{c}{\omega}\int d\omega S(\omega)\mathfrak{J}^{\ell m_\ell m_{ph}}(\mu\omega/cM_e;\kappa_O:\kappa_F)\mathcal{C}^{(b;\kappa)}_{\ell,m_\ell,m_{ph};\Delta m_R,\mu,\alpha_\rho,\alpha_z}
 \end{equation}
 the parameters $m_{ph}$, $\ell$ and $m_\ell$ are directly connected to the internal atom transitions as described in Eq.~(\ref{eq:Jplus}), that is $m_{ph}$ keeps the information about the vectorial coupling of the electron current to the EM field, $\ell$ = 0 for internal dipole transitions, and $\ell$ =1, $m_\ell$=0,$\pm 1$ for electric quadrupole and magnetic dipole transitions.
Higher multipole internal atomic transitions can be evaluated taking into account that, in terms of $k_\bot$ and $\varphi_{\mathbf{k}_\bot}$,
\begin{equation}
Y_{\ell,m_\ell}(\hat{\mathbf{k}}) = N_{\ell,m_\ell}
e^{im_\ell\varphi_{\mathbf{k}_\bot}}{\mathcal{P}}_\ell^{m_\ell}
(\sqrt{1-(k_\bot c/\omega)^ 2})
\end{equation}
with ${\mathcal{P}}_\ell^{m_\ell}$ the associated Legendre functions.
 The parameters $\Delta m_R$, $\mu$, $\alpha_\rho$,and $\alpha_\rho$ in Eq.~(\ref{eq:cbmk}), encode the information of the center of mass transition matrix elements,
Eq.~(\ref{eq:wfcc}-\ref{eq:wfcc4}) in Appendix C. The functions $\mathcal{R}$ and ${\mathcal{Z}}$ exhibit an exponential factor \begin{equation}\mathcal{R}{\mathcal{Z}}\sim e^{-\omega^2\alpha_z^2/4c^2}e^{-k_\bot^2(\alpha^2_\bot-\alpha^2_z)/4}\label{eq:exp}.\end{equation}  For a  spherical symmetric trap the equivalent exponential factor is recovered, Eq.~(\ref{eq:ss1}) Appendix C. The electromagnetic field structure is determined by the polarization vector ${\mathbf{e}^{(b)}_\kappa}$ and the angular spectrum $\tilde \Psi_{\kappa}^{(b)}(\mathbf{k}_\bot)$.

For Bessel modes the quantum numbers are the polarization $b$, the transverse wave number $\kappa_\bot$ and the topological charge $m_{B}$.  The evaluation of the integrals is direct, for scalar beams
\begin{equation}
{\mathcal{C}}^{(\pm;m_{B},\kappa_\bot)}_{m_{ph},\ell,m_\ell;\Delta m_R,\mu,\alpha_\rho,\alpha_z} =
\mathcal{N}^{sc}_{\kappa_\bot,\omega}
\big[\delta_{m_{ph},\pm 1}\delta_{m_\ell,-m_{B}}- \frac{\kappa_\bot}{\sqrt{1-(\kappa_\bot c/\omega)^ 2}}\delta_{m_{ph},0}\delta_{m_\ell,m_{B}\pm 1}\big]\mathbb{C}^{\ell,m_\ell}_{\Delta m_R,\mu,\alpha_\rho,\alpha_z}(\kappa_\bot)\label{eq:BBcs1}
\end{equation}
while for vector beams
\begin{eqnarray}
\mathcal{C}^{(TE;m_{B},\kappa_\bot)}_{\ell,m_\ell;\Delta m_R,\mu,\alpha_\rho,\alpha_z}  =\mathcal{N}^{TE}_{\kappa_\bot,\omega} \frac{\kappa_\bot c}{\sqrt{2}\omega i}&\big[&\delta_{m_{ph}, 1}\delta_{m_\ell, 1-m_{B}} -\delta_{m_{ph}, -1}\delta_{-m_\ell,m_{B}+1}\big]\mathbb{C}^{\ell,m_\ell}_{\Delta m_R,\mu,\alpha_\rho,\alpha_z}(\kappa_\bot)\label{eq:BBcs2}
\\
\mathcal{C}^{(TM;m_{B},\kappa_\bot)}_{\ell,m_\ell;\Delta m_R,\mu,\alpha_\rho,\alpha_z}= \mathcal{N}^{TM}_{\kappa_\bot,\omega}\Big[ \frac{\kappa_\bot c\sqrt{1-\kappa_\bot^2c^2/\omega^2}}{\sqrt{2}\omega i}&\big[&\delta_{m_{ph}, 1}\delta_{m_\ell,1-m_{B}} +\delta_{m_{ph}, -1}\delta_{-m_\ell,m_{B}+1}\big]\nonumber\\ &-&\frac{\kappa_\bot^2\omega^2}{c^2}\delta_{m_\ell,-m_{B}}\Big]
\mathbb{C}^{\ell,m_\ell}_{\Delta m_R,\mu,\alpha_\rho,\alpha_z}(\kappa_\bot)\label{eq:BBcs3}
\end{eqnarray}
with \begin{equation}\mathbb{C}^{\ell,m_\ell}_{\Delta m_R,\mu,\alpha_\rho,\alpha_z}(\kappa_\bot)=2\pi N_{\ell,m_\ell}
{\mathcal{R}}^{\Delta m_{R},\mu}({\kappa_\bot\alpha_\rho}){\mathcal{Z}}^{n_z,n^\prime_z}(\alpha_z\sqrt{(\omega^2/c^2-\kappa_\bot^2}){\mathcal{P}}_\ell^{m_\ell}
(\sqrt{1-(\kappa_\bot c/\omega)^ 2}).\end{equation}
These analytical expressions  allow an optimization procedure of the parameters to achieve enhancements or inhibitions of the transitions rates. This is directly observed from the delta factors which determine the values of the Bessel beam topological charge for a given beam polarization that yield specific changes in the atomic
quantum number; but it also applies for the selection of the $\kappa_\bot$ parameter as illustrated below.

 Laguerre- Gaussian (LG) modes, usually studied in the paraxial regime \cite{Davila,alexandrescu} have an angular spectra given by Eq.~(\ref{def:laggauss:as}).
 The vortex factor $(k_\bot w_0/c)^me^{im\varphi_{\mathbf{k}_\bot}}$  leads to similarities and important differences between the integrals related to  Laguerre Gaussian and to Bessel modes: while both share the same value of the integration on the $\varphi_{\mathbf{k}_\bot}$ variable, LG beams become zero as $k_\bot\rightarrow 0$ in a continuous way while for Bessel beams the variable $k_\bot$ takes abruptly  the nonzero value $\kappa_\bot$. The angular similarity has as a consequence that both Bessel and LG beams can be used to enhance or inhibit transitions involving selected changes in the atomic (CM and internal) angular momentum as observed in Eq.~(\ref{eq:BBcs1}-\ref{eq:BBcs3}), and mentioned in the last section for CM motion with spherical symmetry. A detailed analysis on such a direction is reported, neglecting retardation effects, for Bessel beams in Ref.~\cite{jauregui} and for Laguerre Gaussian beams in  Ref.~\cite{alexandrescu}.  In the latter case, the interaction electric Hamiltonian of the PZW formalism is used, free motion is assumed for the center of mass in the $z$ direction, and the particular case $p=0$ of the paraxial Laguerre Gaussian modes  are studied. A quite interesting  addition theorem for the factorization of the center or mass and the internal motion adequate for cylindrical symmetry is also introduced in Ref.~\cite{alexandrescu} . The allowed atomic and center of mass transitions for a given polarization and topological charge of the beams as described by Eqs.(\ref{eq:BBcs1}-\ref{eq:BBcs3}) is consistent with the reported results in both references when the appropriate limits are taken.

Let us illustrate an analytical procedure that can be used to optimize the selection of the $\omega_o$ (LG beams) and $\kappa_\bot$ (Bessel beams) parameters to enhance a given transition rate. LG transitions are determined by the elementary factors,

\begin{equation}
e^{-\omega^2\alpha_z^2/4c^2}\int_0^1 e^{-x^2w_{LG}^2/4}  (\sqrt{1-x^2})^sx^rxdx,\quad w_{LG}^2 = (\alpha^2_\rho - \alpha^2_z + w_o^2)\omega^2/c^2,
\end{equation}
The function $ e^{-x^2w_{LG}^2/4}  (\sqrt{1-x^2})^sx^r$
becomes zero at $x=0$, 1, if $s$, $r$ $\ne 0$, and the factor $(\sqrt{1-x^2})^sx^r$ achieves its maximum value at $x_m^2=r/(r+s)<1$. If $r=0$ and $s\ne 0$ (or  $s=0$ and $r\ne 0$)  the maximum is achieved at $x_m = 0$ (or  $x_m =1$). As a consequence
 $$
\int_0^1 e^{-x^2w_{LG}^2/4}  (\sqrt{1-x^2})^sx^rxdx < (1-x_m^2)^{s/2}x_m^{r/2}\int_0^1 e^{-x^2w_{LG}^2/4}x dx
=\frac{(1-x_m^2)^{s/2}x_m^{r/2}}{2}\frac{1-e^{-w_{LG}^2/4}}{w_{LG}^2/4}
$$
and
\begin{equation}
e^{-\omega^2\alpha_z^2/4c^2}\int_0^1 e^{-x^2w_{LG}^2/4}(\sqrt{1-x^2})^sx^rxdx<
\frac{(1-x_m^2)^{s/2}x_m^{r/2}}{2}\frac{e^{-\alpha_z^2\omega^2/4c^2}-e^{-(\alpha_\rho^2+ w_o^2)\omega^2/4c^2}}{(\alpha^2_\rho - \alpha^2_z + w_o^2)\omega^2/4c^2}
\end{equation}
The maximum value of the latter term is one, and it is achieved for an elongated trap with $\alpha^2_z = \alpha^2_\rho + w_o^2\rightarrow 0$, exhibiting and exponential decay on any direction of the $(\alpha_z,\sqrt{\alpha^2_\rho + w_o^2})$ plane, with independence on the values of the positive parameters $r$ and $s$.
As a consequence, transition rates are enhanced if the Laguerre Gauss beam is highly focused ($w_o \omega/c$ as small possible)  and the trap is an elongated one with $\alpha^2_z = \alpha^2_\rho + w_o^2$ as small as possible.  This is consistent with earlier observations \cite{jauregui} that predict that atomic transitions can be enhanced by locating the atom close to a light vortex. Note that for $\alpha_\rho \ll 1$, working with paraxial LG beams
yields
$$ e^{-\omega^2\alpha_z^2/4c^2}\int_0^1 e^{-x^2w_{LG}^2/4}(\sqrt{1-x^2})^sx^rxdx < \frac{(1-x_m^2)^{s/2}x_m^{r/2}}{2}\frac{e^{-\alpha_z^2\omega^2/4c^2}}{w_o^2\omega^2/4c^2}, \quad w_o^2\omega^2/4c^2 \gg 1$$
with the last factor being much smaller than one.

For Bessel beams, and given values of the parameters $\alpha_{z,\rho}$, the external parameter $\kappa_\bot c/\omega$ can be chosen to achieve the maximum value of the modulus of $\mathcal{C}$. Calculations can be performed in an analytical way for the elementary term
\begin{equation}
e^{-\omega^2\alpha_z^2/4c^2} e^{-x_o^2w_{B}^2/4}  (\sqrt{1-x_o^2})^sx_o^r,\quad x_o =\kappa_\bot c/\omega,\quad  w_{B}^2 = (\alpha^2_\rho - \alpha^2_z)\omega^2/c^2,
\end{equation}
which is also zero at $x_o = 0$ and $x_o=1$ and achieves an extreme value for $x_o$ given by
\begin{equation}
\frac{\kappa_\bot^2 c^2}{\omega^2} = \frac{w_{B}^2/2 + s + r \mp \sqrt{(w_{B}^2/2 + s + r)^2 -2 w_B^2r}}{w_B^2},
\end{equation}
the sign is selected according to  $\alpha^2_\rho > \alpha^2_z$ or $\alpha^2_\rho < \alpha^2_z$. If $\alpha^2_\rho = \alpha^2_z$, the maximum is achieved at $\kappa_\bot^2 c^2/\omega^2 = r/(r+s)$. These expressions give the optimal values of $\kappa_\bot$ as a function of the exponents $r$ and $s$, and the geometry of the  trap in terms of the parameter $w_B^2$.

\section{Conclusions.}

In order to design experiments towards the control of atomic transition rates via laser light shaping, it is necessary to perform
quantitative equitable calculations that allow the identification of the optimal parameters to inhibit or enhance such rates with respect to  standard
conditions where, in most cases, Gaussian laser beams in the paraxial regime are used.

In this work we have shown that these calculations can be easily performed whenever a prior identification of the internal and center of mass states of interest is done; this follows from the fact that
the structure of radiative atomic  transition amplitudes can always be written in terms of:
\begin{itemize}
\item[(a)] An internal atomic transition factor.  It has a vectorial character for the dominant
interaction Hamiltonian $H_{I1}$ that results from the coupling of the electron current density to the electromagnetic field, and a scalar character for other terms in the non relativistic interaction Hamiltonian including one term in the coupling of the nucleus current density to the electromagnetic field, and the terms quadratic in the electromagnetic field.
The vectorial factor  $\mathbf{J}^{\pm}_{rel}$ can be evaluated using different approaches to the description of the atomic dynamical space, for instance, a non relativistic approach as that implicit in Eq.~(\ref{eq:trans}) or a fully relativistic approach including the electron spin as the one given in Ref.~\cite{landau}. The central field structure of atoms makes that the dependence of the internal atomic transition factors on the electromagnetic wave vector $\mathbf{k}$ can be naturally  expressed in terms of spherical harmonics $Y_{\ell,m}(\hat{\mathbf{k}})$  in all cases. 

\item[(b)] An atomic center of mass  transition factor  that describes elementary recoil effects. Under standard conditions, the velocities $P/M_T$ involved in the CM motion are much smaller than those associated to the internal motion $p_{rel}/\mu$; that reflects on smaller values of the matrix elements of the interaction Hamiltonian that couples the electromagnetic field to the nucleus current density  than those related to the EM field-electron current density coupling. The atomic center of mass  transition factor  has a scalar character for the dominant interaction Hamiltonian $H_{I1}$  and for the terms quadratic in the electromagnetic field, and a vectorial character for the other terms in the minimal coupling interaction Hamiltonian. The vectorial factor $\mathbf{I}_{CM}$  was shown to correspond to the gradient in wave vector space of its scalar analogue ${I}_{CM}$. The center of mass transition factor can play an active role in the control of internal atomic transition rates. This has already been recognized in Ref.~\cite{jauregui}, where it was shown that the probability that the internal state of an atom acquires orbital angular momentum from  a Bessel mode is maximum for an atom located close to the symmetry axis of the beam. For a center of mass Gaussian packet, the relevant parameter was found to be the ratio of the spread of the wave packet to the transversal wavelength of the photon. In the present formalism this corresponds to taking $\mathbf{R}_0 =\mathbf{0}$ in Eq.~(\ref{eq:icm}) and
$\hbar\omega^2/M_T\omega_{CM}c^2\sim 1$ to maximize the ${\mathcal{I}}_{CM}(\hbar\omega^2/M_T\omega_{CM}c^2)$ value
in Eq.~(\ref{eq:cmYlma}). In the case of an idealized CM description in terms of plane waves the calculation of the elementary recoil effects is direct. The CM transition factors  are necessary to perform, via their incorporation in the corresponding Bloch equations,  realistic simulations of experiments at room temperature and moderate densities.

\item[(c)] A factor that involves the EM radiation structure which is coupled to the  internal transitions and
recoil terms.  The EM radiation is characterized in this term by its angular spectra $\tilde{\Psi}_{\kappa;\omega}^{(b)}(k_x,k_y)$
and its polarization which, in general, also  depends on $\mathbf{k}$.
\end{itemize}

The electromagnetic structure is completely encoded in the factor (c) that can be optimized to enhance or inhibit the
transitions of interest.  The general procedure has been exemplified for atoms with center of mass available states described by (i) plane waves, (ii) wave functions
exhibiting spherical symmetry and (iii) wave functions for traps with cylindrical symmetry.
In the second case,  EM shaping is  determined by the integrals ${\mathcal{H}}^{(\kappa_{EM})}_{\ell,m_\ell}$ given by Eq.~(\ref{eq:hbmk}); a quantitative comparative analysis between the
transition rates for different EM fields was shown to be direct, as illustrated  for spherical waves, Gaussian beams with different waists and propagation invariant
beams with diverse angular spectra.

Spherical modes provide an ideal scenario for the description of spontaneous atomic transitions between internal states.  Their usage
for the control of induced transitions is conditioned by the capability of creating  spherical modes
with predetermined characteristics. An upper bound for the important functions ${\mathcal{H}}^{(\kappa_{EM})}_{\ell,m_\ell}$ for spherical waves was found, Eq.~(\ref{eq:sphh}). This value is also an upper bound for most of the corresponding integrals obtained for Gaussian (Fig.~1) and propagation invariant modes (Fig.~2). An interesting exception corresponds to transverse magnetic Gaussian beams inducing $\pi$ transitions ($\hat{\mathbf{e}}_{ph} =\hat{\mathbf{e}}_{0}=\hat{\mathbf{e}}_{z})$ in the antiparaxial regime, Fig.~2b.

This study also allowed the recognition of the interactive role of polarization and EM angular spectra.
For electric quadrupole transitions, induced by $H_{I1}$, the  selection rule $\Delta J =0, \pm 1, \pm 2$  allows changes of $2\hbar$
 in the atomic angular momentum. This yields an implicit requirement of light modes with orbital angular momentum.
 These modes can be spherical (as worked out in Section VII.A) or with other symmetries like Bessel modes or Laguerre Gauss
  modes. The plane waves decomposition in terms of spherical modes, Eq.~(\ref{eq:sph}), makes explicit their
  orbital angular momentum content.

By changing the ratio between the transverse and longitudinal components in the dominant wave vectors
  yielding a structured beam, the relevance of predetermined internal states involved in a forbidden transitions can be modified.
  This is a direct physical consequence of the behavior of the functions ${\mathcal{H}}^{(\kappa_{EM})}_{\ell,m_\ell}$ that
  encode the polarization  and angular spectra effects on transition rates.
 By evaluating these functions,  it was shown that,
 in the paraxial limit and for scalar Gaussian beams,  any multipole rate depends on the EM
 energy per unit transverse area. We also noticed that measurable increments of the transition
 rates for Gaussian beams with respect to this asymptotic value requires waists of the order
 of the wavelength. This is not a trivial task and, since Gaussian  modes
 (as well as Laguerre-Gauss and Hermite-Gauss beams) have a transverse
wave vector  distribution centered around $\mathbf{k}_\bot = \mathbf{0}$,
 they are not necessarily the optimal choice for achieving that enhancement.

  For propagation
 invariant modes, we have noticed that they could be used to enhance transition rates in more favorable conditions
 since their approximate implementation involves a narrow wave vector  distribution centered around $\mathbf{k}_\bot \ne \mathbf{0}$.

The relevance of the angular spectrum on recoil effects was studied in general and illustrated for Mathieu, Bessel and Weber beams.
Using a plane wave description of the center of mass motion, we  showed that the EM angular spectrum can be directly
imprinted in the atomic recoil effects for dipole transitions. In particular, our calculations confirm that, via dipole transitions,
photons with orbital angular momentum can transfer this dynamical variable to the center of mass \cite{Davila,jauregui,alexandrescu}. For
electric quadrupole and magnetic dipole transitions, the angular spectrum and polarization have diverse and
interesting consequences on internal and center of mass transition rates. For those transitions, the
effective interaction term describing recoil effects depends directly on the gradient of the EM field, so that semiclassical forces could depend on second derivatives of the electric and magnetic fields. It was also shown that Mathieu beams can be used to entangle atomic internal and center of mass degrees of freedom when the available states of the latter have spherical symmetry.

Another important observation is that the formalism developed in this work can be adapted to other atomic systems such as molecules or nanostructures.
In such cases, the particle that describes the transition of interest is not necessarily an electron. One could be interested, for instance, on the rotation or vibration of  atomic nuclei. In any case, the symmetry of the internal available states should be taken into account for the most efficient calculation of the elementary integral containing the EM factor $e^{i\mathbf{k}\cdot\mathbf{r}}$.

 If the atoms of interest are embedded  in a material media,  the electromagnetic field can be described using an equation similar to Eq.~(\ref{eq:EF}) by replacing the proper dispersion relation (which may depend on the polarization). Evanescent waves would also require a modification on the Fourier factor $ e^{i{\mathbf{k}}\cdot{\mathbf{r}}}$ to incorporate an exponential decay in the adequate direction. Then, the general formalism presented in this work would be properly adapted to describe transition rates in those systems.

\vspace{1cm}

\acknowledgements{I thank S. Hacyan and R. Guti\'errez-J\'auregui for their useful comments.  This
work was partially funded by the  grants CONACYT  166961 and LN-232652.} 

\vspace{1cm}

\appendix{\bf Appendix A. Normalization of the EM modes}

The quantization of the electromagnetic field in terms of modes is usually based on the orthonormality  requirement
\begin{equation}
\frac{1}{8\pi}\int d^3\mathbf{r} \Big(\mathbf{E}_\kappa^{(b)*}(\mathbf{r})\cdot  \mathbf{E}_{\kappa^\prime}^{(b)}(\mathbf{r}) +\mathbf{B}_\kappa^{(b)*}(\mathbf{r})\cdot  \mathbf{B}_{\kappa^\prime}^{(b)}(\mathbf{r})\Big) = \hbar\omega \delta(\omega - \omega^\prime) \delta_{\kappa,\kappa^\prime},\label{eq:ortho}
\end{equation}
with the integration performed over the whole space ${\mathfrak{R}}^3$.
For modes with polarization $TE$, $TM$  this  condition has a simple expression in terms of an integral of the modulus of the angular spectra,
\begin{equation}
2\pi^2\int d^2k_\bot\frac{k_z c}{\omega}\frac{k_\bot^2 c^2}{\omega^2}\tilde\Psi^*_{\kappa^\prime}(\mathbf{k}_\bot)\Psi_\kappa(\mathbf{k}_\bot) = \hbar\omega \delta_{\kappa,\kappa^\prime}.\label{eq:TETM}
\end{equation}
That is not the case for structured scalar modes due to the fact that the polarization vectors $\mathfrak{E}^{(1)}$ and  $\mathfrak{E}^{(2)}$ are not orthogonal. This kind of modes are frequently used in experimental realizations where
circularly polarized Gaussian beams impinge atomic samples.
For scalar localized modes, Eq.~(\ref{eq:TETM}) can be taken just as a normalization condition,
\begin{equation}
\frac{1}{8\pi}\int  \Big(\mathbf{E}_\kappa^{(\pm)*}\cdot  \mathbf{E}_{\kappa}^{(\pm)} +\mathbf{B}_\kappa^{(\pm)*}\cdot  \mathbf{B}_{\kappa}^{(\pm)}\Big) = \pi^2\int d^2k_\bot\frac{k_z c}{\omega}\Big[1 + \frac{\omega^2}{k_z^2 c^2}\Big]\vert\tilde\Psi_{\kappa^\prime}(\mathbf{k}_\bot)\vert^2.\label{eq:circ2}
\end{equation}
In both Eqs.~(\ref{eq:TETM}-\ref{eq:circ2}), the integral in the left side must be performed within the circle of radius $\omega/c$ in the $\mathbf{k}_\bot$ space.
For axial symmetric scalar Gaussian modes of waist $\mathrm{w}_\bot$,
$$\tilde \Psi_{\mathrm{w}_\bot}(\mathbf{k}_\bot) = {\mathcal{N}_{\mathrm{w}_\bot}} \mathrm{e}^{-\mathrm{w}_\bot^2 k_\bot^2},$$
and circular polarization, the exact expression of the  normalization integral valid beyond the paraxial approximation is
 \begin{eqnarray}
 \frac{1}{8\pi}\int  \Big(\mathbf{E}_\kappa^{(\pm)*}\cdot  \mathbf{E}_{\kappa^\prime}^{(\pm)} +\mathbf{B}_\kappa^{(\pm)*}\cdot  \mathbf{B}_{\kappa^\prime}^{(\pm)}\Big) & = & \vert{\mathcal{N}_{\mathrm{w}_\bot}}\vert^2\frac{\omega^2\pi^3}{c^2}\Big[\frac{1}{2a^2} +\sqrt{2\pi}\frac{4a^2 -1}{8 a^3}e^{-2a^2} \mathrm{Erfi}(\sqrt{2}a)\Big] \nonumber \\
   & = &\hbar\omega,\end{eqnarray}
   with $a =\mathrm{w}_\bot\omega/c$ and Erfi the imaginary error function.
In the case of Hermite-Gaussian modes of waists $\mathrm{w}_{x,y}$ that guarantee the validity of the paraxial approximation, $\vert k_z\vert \sim\omega/c$ the polarization vectors $\mathfrak{E}^{(1)}$ and $\mathfrak{E}^{(2)}$ are quasi orthogonal, and the integral over the  circle of radius $\omega/c$ can be replaced to that over the whole space. Then the orthonormalization condition is approximately achieved by taking ${\mathcal{N}} = \sqrt{\mathrm{w}_x\mathrm{w}_y\hbar\omega/2^{m+n}\pi^3 m! n!}$.

Propagation invariant beams are non localized waves  (the electric fields do not become zero fast enough as $\vert{\mathbf{x}}\vert\rightarrow \infty$)so that the  orthonormalization requirement Eq.~(\ref{eq:ortho}) gives rise to
delta functions on the modulus of $k_\bot$. Explicit results for TE and TM modes have been reported for Bessel\cite{shj}, Mathieu\cite{brlrjm} and Weber\cite{brlrjw} modes.

 An alternative for non localized waves, corresponds to evaluate normalization factors within a finite space region
\begin{equation}
\frac{1}{8\pi}\int_\mathcal{V}  \Big(\mathbf{E}_\kappa^{(b)*}\cdot  \mathbf{E}_{\kappa}^{(b)} +\mathbf{B}_\kappa^{(b)*}\cdot  \mathbf{B}_{\kappa^\prime}^{(b)}\Big) = \hbar\omega.
\end{equation}
For plane waves this normalization usually considers a parallelepiped
of volume $\mathcal{V}=L_1L_2L_3$.
This method has the advantage of being more intuitive and closer to experimental realizations.  Taking this into account, we outline
the main steps of this procedure for other propagation invariant beams. In that case, the natural space region for performing the
normalization condition is a cylinder with symmetry axis along the main propagation axis, radius $R$ and length $L$.
The normalization factor should then be written in terms of the angular spectra given above.
A relevant result is:
\begin{equation}
\int_{\mathcal{V}_{R,L}} d^3 r e^{i(\mathbf{k}^\prime-\mathbf{k})\cdot \mathbf{r}} = 2\pi R^2 L\Big[\frac{J_1(\vert\mathbf{k}_\bot^\prime -\mathbf{k}_\bot\vert R)}{\vert\mathbf{k}_\bot^\prime -\mathbf{k}_\bot\vert R}\Big]\Big[\frac{\sin(k_z^\prime -k_z)L/2}{(k_z^\prime -k_z)L/2}\Big]
\end{equation}
Then, given a function of the form
\begin{equation}
\Phi_{\kappa}(\mathbf{r}) = \int d^2\mathbf{k}_\bot \phi_{\kappa}(\mathbf{k}_\bot) e^{i\mathbf{k}\cdot\mathbf{r}},\quad\quad k_z = \sqrt{\frac{\omega^2}{c^2}-k_\bot ^2},
\end{equation}

\begin{equation}
\int_{\mathcal{V}_{R,L}}d^3r\Phi^*_{\kappa^\prime}(\mathbf{r})\Phi_\kappa(\mathbf{r}) = 2\pi R^2 L\int d^2\mathbf{k}_\bot^\prime d^2\mathbf{k}_\bot\phi^*_{\kappa^\prime}(\mathbf{k}^\prime_\bot)\phi_\kappa(\mathbf{k}_\bot)\Big[\frac{J_1(\vert\mathbf{k}_\bot^\prime -\mathbf{k}_\bot\vert R)}{\vert\mathbf{k}_\bot^\prime -\mathbf{k}_\bot\vert R}\Big]\Big[\frac{\sin(k_z^\prime -k_z)L/2}{(k_z^\prime -k_z)L/2}\Big].
\end{equation}
For $L>>c/\omega$, the last term is non negligible just for $k_z^\prime \sim k_z$ and, if we consider two modes with the same frequency, necessarily
$\vert\mathbf{k}_\bot\vert \sim \vert\mathbf{k}_\bot^\prime\vert$.
For propagation invariant TE or TM beams, the orthonormalization integral becomes
$$
\frac{1}{8\pi}\int_\mathcal{V}  \Big(\mathbf{E}_\kappa^{(b)*}\cdot  \mathbf{E}_{\kappa}^{(b)} +\mathbf{B}_\kappa^{(b)*}\cdot  \mathbf{B}_{\kappa^\prime}^{(b)}\Big) = $$
\begin{equation}
2\pi R^2 L \frac{\kappa_\bot^2c^2}{4\pi\omega^2}\int d\varphi_{\mathbf{k}^\prime}\int d\varphi_{\mathbf{k}}
f_{\kappa^\prime}^*(\varphi_{\mathbf{k}^\prime})f(\varphi_{\mathbf{k}})\frac{J_1(\sqrt{2}\kappa_\bot R\sqrt{1-\cos(\varphi_{\mathbf{k}^\prime}-\varphi_{\mathbf{k}})})}
{\sqrt{2}\kappa_\bot R\sqrt{1-\cos(\varphi_{\mathbf{k}^\prime}-\varphi_{\mathbf{k}})}}.
\end{equation}
For $R>>\kappa_\bot^{-1}$,
\begin{equation}
 2R \frac{J_1(\sqrt{2}\kappa_\bot R\sqrt{1-\cos(\varphi_{\mathbf{k}^\prime}-\varphi_{\mathbf{k}})})}
{\sqrt{2}\kappa_\bot R\sqrt{1-\cos(\varphi_{\mathbf{k}^\prime}-\varphi_{\mathbf{k}})}}\rightarrow \delta (\sqrt{2}\kappa_\bot \sqrt{1-\cos(\varphi_{\mathbf{k}^\prime}-\varphi_{\mathbf{k}})}).
\end{equation}
So that,
\begin{eqnarray}
\frac{1}{8\pi}\int_\mathcal{V}  \Big(\mathbf{E}_\kappa^{(b)*}\cdot  \mathbf{E}_{\kappa}^{(b)} +\mathbf{B}_\kappa^{(b)*}\cdot  \mathbf{B}_{\kappa^\prime}^{(b)}\Big)  &=&\pi R L\frac{\kappa_\bot^2c^2}{4\pi\omega^2}\int d\varphi_{\mathbf{k}^\prime}\int d\varphi_{\mathbf{k}}
f_{\kappa^\prime}^*(\varphi_{\mathbf{k}^\prime})f_{\kappa}(\varphi_{\mathbf{k}})\delta((\sqrt{2}\kappa_\bot \sqrt{1-\cos(\varphi_{\mathbf{k}^\prime}-\varphi_{\mathbf{k}})})\nonumber\\
&=&\frac{\mathcal{V}_{R,L}}{\kappa_\bot R}\frac{\kappa_\bot^2c^2}{4\pi\omega^2} \int d\varphi_{\mathbf{k}}f_{\kappa^\prime}^*(\varphi_{\mathbf{k}})f_{\kappa}(\varphi_{\mathbf{k}}).
\end{eqnarray}
For Bessel beams and Mathieu beams the integral over $\varphi_{\mathbf{k}}$ is simply $2\pi\delta_{m^\prime,m}$. For Weber beams, the parameter $\alpha$ is continuous and
\begin{equation}\int_0^{2\pi} d\varphi_{\mathbf{k}}\frac{e^{i(\alpha-\alpha^\prime) \ln\vert\tan\varphi_{\mathbf{k}}/2\vert}}{4\pi\vert\sin\varphi_{\mathbf{k}}\vert} =\frac{1}{2\pi}\int_{-\infty}^{\infty} e^{i(\alpha-\alpha^\prime)u} = \delta(\alpha-\alpha^\prime).
\end{equation}

An third alternative valid for propagation invariant modes corresponds to defining them in terms of a Gaussian superposition of
plane waves centered at $\kappa_\bot$ instead of the idealized definition given by Eq.~(\ref{def:piof:as}),
\begin{equation}
\tilde{\Psi}^{(b:approx)}_{\mathrm{w}_\bot,\kappa_\bot,\kappa_m}(k_x ,k_y) =\mathcal{N}_{\mathrm{w}_\bot,\kappa_\bot,\kappa_m}^{(b:approx)} \mathrm{e}^{-\mathrm{w}_\bot ^2(k_\bot - \kappa_\bot)^2} \psi^{\kappa_m}(\varphi_{\mathbf{k}}), \quad k_\bot =
    \sqrt{k_x^2 + k_y^2},\quad \varphi_{\mathbf{k}} = \mathrm{atan}(k_y/k_x),
\end{equation}
with $\mathrm{w}_\bot\gg \kappa_\bot$. In the limit $\mathrm{w}_\bot/\kappa_\bot\rightarrow \infty$ with finite $\kappa_\bot$
one recovers the idealized definition.
The elementary integrals necessary to evaluate the normalization factor $\mathcal{N}_{\mathrm{w}_\bot,\kappa_\bot,\kappa_m}^{(b:approx)}$ for Bessel, Mathieu and Weber Gaussian beams are those described in the previous paragraph besides the integral
\begin{equation}
\int_0^\infty  dk_\bot k_\bot \mathrm{e}^{-2\mathrm{w}_\bot ^2(k_\bot - \kappa_\bot)^2} =
\frac{1}{2\mathrm{w}_\bot^2}\int_{\mathrm{w}_\bot\kappa_\bot}^\infty  dt \mathrm{e}^{-2t}=\frac{\mathrm{e}^{- 2\mathrm{w}_\bot\kappa_\bot}}{4\mathrm{w}_\bot^2}.
\end{equation}
\vspace{1cm}

\appendix{\bf Appendix B. Spherical EM waves}
A complete basis for transverse
electromagnetic fields in free space can be constructed in terms of spherical
vectors. In the notation established by Eq.~(\ref{eq:bmodeM}), the corresponding modes are defined by (see, e.
g., \cite{landau})
\begin{equation}
{\mathfrak{E}}^{(b)}_{j m}(k_x ,k_y;\omega) = \frac{\sqrt{\hbar\omega} }{2\pi \vert k_z\vert}\mathbb{Y}_{jm}^{(b)}( \hat{\bf k}), \quad b =E,M,\quad  \hat{\bf k}
= \frac {{\bf k}}{\vert{\bf k}\vert}.
\end{equation}
In these equations, the superscript specifies the electric $(E)$
and magnetic $(B)$ spherical  modes, and
\begin{eqnarray}
\mathbb{Y}^{(E)}_{j{m}}(\theta_{\hat{\bf k}},\varphi_{\hat{\bf k}}) & =&
\frac{1}{j(j+1)}
\nabla_{\hat{\bf k}}Y_{jm}(\theta_{\hat{\bf k}},\varphi_{\hat{\bf k}}),\\
\mathbb{Y}^{(M)}_{jm}(\theta_{\hat{\bf k}},\varphi_{\hat{\bf k}}) &  =&
{\hat{\bf k}}\times \mathbb{Y}^{(E)}_{jm}(\theta_{\hat{\bf k}},\varphi_{\hat{\bf k}}),
\end{eqnarray}
with
\begin{equation}
\nabla_{\hat{\bf k}}=\hat {\bf \theta}_{\bf
k}\frac{\partial}{\partial\theta_{\bf k}}+\hat{\bf \varphi}_{\bf
k}\frac{1}{\sin\theta_{\bf k}}\frac{\partial}{\varphi_{\bf k}};
\end{equation}
$Y_{jm}(\theta_{\hat{\bf k}},\varphi_{\hat{\bf k}})$ are the
spherical harmonics, and
\begin{equation}
\hat{\bf \theta}_{\bf k}= \cos\theta_{\bf k}\cos\varphi_{\bf k}
\hat{\bf e}_1+\cos\theta_{\bf k}\sin\varphi_{\bf k} \hat{\bf e}_2
-\sin\theta_{\bf k} \hat{\bf e}_3
\end{equation}
and
\begin{equation}
\hat{\bf \varphi}_{\bf k}= -\sin\varphi_{\bf k}\hat{\bf e}_1
+\cos\varphi_{\bf k}\hat{\bf e}_2.
\end{equation}
It can be shown \cite{landau} that,
\begin{equation}
\mathbb{Y}^{(b)}_{j{m}}= \sum_{\lambda=0,\pm1} (\mathbb{Y}^{(b)}_{j{m}})_\lambda \hat{e}_\lambda, \quad b=E,M
\end{equation}
\begin{eqnarray}
(-1)^{j+m+\lambda+1} (\mathbb{Y}^{(E)}_{j m})_\lambda = &-&c_\lambda\sqrt{j}
\begin{pmatrix}j+1 & 1 & j\cr m+\lambda &-\lambda &-m\end{pmatrix}Y_{j+1,m+\lambda} \nonumber\\
&+&c_\lambda\sqrt{j+1}\begin{pmatrix}j-1&1&j\cr m+\lambda &-\lambda &-m \end{pmatrix}Y_{j-1,m+\lambda}\nonumber\\
(-1)^{j+m+\lambda+1} (\mathbb{Y}^{(M)}_{j m})_\lambda = &-&c_\lambda\sqrt{2j+1}\begin{pmatrix}j&1&j\cr m+\lambda &-\lambda &-m\end{pmatrix}Y_{j,m+\lambda}
\end{eqnarray}
with $c_0 = i$, $c_+ = -i$  and $c_- = i$. From these expressions an explicit relation for the basic integrals
$\vert {\mathcal{H}}^{(E;\lambda;jm)}_{\ell,m_\ell}\vert$ defined in Eq.~(\ref{eq:sphH}) can be found,

$$\vert {\mathcal{H}}^{(E;\lambda;jm)}_{\ell,m_\ell}\vert =\frac{\sqrt{\hbar\omega^3}}{2\pi c}\cdot\delta_{m_\ell,m+\lambda}$$
\begin{eqnarray}
\cdot \Big[ \delta_{\ell,j+1}\sqrt{\frac{\ell-1}{\ell}}\Big[\delta_{\lambda,0}\sqrt{\frac{(\ell-m_\ell)(\ell +m_\ell)}{4\ell^2-1}}
&+&\delta_{\lambda,1}\sqrt{\frac{(\ell+m_\ell-1)(\ell +m_\ell)}{2(4\ell^2-1)}}+\delta_{\lambda,-1}\sqrt{\frac{(\ell-m_\ell -1)(\ell -m_\ell)}{2(4\ell^2-1)}}\Big]\nonumber \\
+ \delta_{\ell,j-1}\sqrt{\frac{\ell+2}{\ell +1}}\Big[\delta_{\lambda,0}\sqrt{\frac{(\ell -m_\ell+1)(\ell +m_\ell+1)}{4(\ell +1)^2-1}}
&+&\delta_{\lambda,1}\sqrt{\frac{(\ell-m_\ell)(\ell-m_\ell +1)}{2(4(\ell +1)^2-1)}}+\delta_{\lambda,-1}\sqrt{\frac{(\ell+m_\ell+2)(\ell+m_\ell +1)}{2(4(\ell +1)^2-1)}}\Big]\nonumber
\end{eqnarray}

\begin{equation}
\vert {\mathcal{H}}^{(M;\lambda;jm)}_{\ell,m_\ell}\vert =\frac{\sqrt{\hbar\omega^3}}{2\pi c}\cdot\delta_{m_\ell,m+\lambda}\delta_{\ell,j}\sqrt{\frac{1}{\ell(\ell+1)}}\Big[\delta_{\lambda,0}\vert m_\ell\vert +\delta_{\lambda,1}\sqrt{\frac{(\ell-m_\ell +1)(\ell+m_\ell)}{2}}+\delta_{\lambda,-1}\sqrt{\frac{(\ell-m_\ell -1)(\ell+m_\ell+2)}{2}}\Big]
\end{equation}

\appendix{\bf Appendix C. Transition center of mass integrals for trapped atoms.}

For the particular case of
a confining potential that has cylindrical symmetry around the $Z$-axis we write
\begin{equation}
\Phi_{n_{\rho},m_{CM},n_z}({\bf R}) =\frac{1}{\sqrt{2^n n!\alpha_z}}\frac{\sqrt{2  p!}}{\sqrt{\alpha_\rho^2(p +\vert m_{CM} \vert)!}}
e^{-\rho^2/2\alpha_\rho^2}e^{-z^2/2\alpha_z^2}(\rho/\alpha_\rho)^{\vert m_{CM}\vert}L_p^{\vert m_{CM}\vert}(\rho^2/\alpha_\rho^2)H_{n_z}( z/{\alpha_z}) e^{im_{CM}\varphi_R}\label{eq:wfcc}
\end{equation}
with $\alpha_{z,\rho}= \sqrt{\pi\hbar/M_T\omega_{z,\rho}}$, and $p = (n_{\rho} + \vert m_{CM}\vert)/2$. The relevant integrals are
\begin{equation}
\int_0^{2\pi} d\varphi_R e^{ik_\bot \rho\cos\varphi_R} e^{i\Delta m_{R}\varphi_R} = 2\pi i^{\Delta m_{R}} J_{\Delta m_{R}}(k_\bot \rho)\label{eq:wfcc1}
\end{equation}
$$\int_0^\infty d\rho J_{\Delta m_{R}}(k_\bot \rho)e^{-\rho^2/\alpha^2_\rho} \rho^{\mu -1} =$$
\begin{equation}
 \Gamma\Big( \frac{\Delta m_R + \mu}{2}\Big)
e^{- k_\bot^2\alpha^2_\rho/4}\Big(\frac{k_\bot^{\Delta m_R}\alpha^{\Delta m_R +\mu}_\rho}{2^{\Delta m_R+1}}\Big)
M((\Delta m_R -\mu)/2 +1, \Delta m_R +1;\frac{k_\bot^2\alpha^2_\rho}{4}),\label{eq:wfcc2}
\end{equation}
and $M$ the confluent hypergeometric function. In particular
\begin{equation}
\int_0^\infty dR_\bot J_{\Delta m_{R}}(k_\bot R_\bot)e^{-R_\bot^2/\alpha^2_\rho} R_\bot^{\Delta m_{R} +1}
=e^{- k_\bot^2\alpha^2_\rho/4}\frac{k_\bot^{\Delta m_R}\alpha^{2\Delta m_R +2}_\rho}{2^{\Delta m_R+1}}.\label{eq:rwfcc3}
\end{equation}

It results convenient to define
$${\mathcal{R}}^{\Delta m_{R},\mu}({k_\bot\alpha_\rho}) = \int_0^{2\pi} d\varphi_R\int_0^\infty dR_\bot e^{i{\mathbf{k}}_\bot\cdot{\mathbf{R}}_\bot}e^{i\Delta m_R\varphi_R}e^{-R_\bot^2/\alpha^2_\rho}R_\bot^{\mu -1}$$
$$= 2\pi i^{\Delta m_R}\Gamma\Big( \frac{\Delta m_R + \mu}{2}\Big)$$
$$e^{- k_\bot^2\alpha^2_\rho/4}\Big(\frac{k_\bot^{\Delta m_R}\alpha^{\Delta m_R +\mu}_\rho}{2^{\Delta m_R+1}}\Big)
M((\Delta m_R -\mu)/2 +1, \Delta m_R +1;\frac{k_\bot^2\alpha^2_\rho}{4}),$$

$${\mathcal{Z}}^{n_z,n^\prime_z}(k_z\alpha_z)=\langle n_z\vert e^{i(k_z\alpha_z)(\hat a + \hat a^\dagger)}\vert n_z^\prime\rangle = $$
\begin{equation}
\Big(ik_z\alpha_z\Big)^{n_z+n^\prime_z} \sum_{\eta=0}^{min\{n_z,n^\prime_z\}}
\sqrt{\frac{n_z!}{(n_z -\eta) !}}\sqrt{\frac{n_z^\prime !}{(n^\prime_z-\eta)!}
}\frac{(ik_z\alpha_z)^{-2\eta}}{\eta !}.\label{eq:wfcc4}
\end{equation}

For the particular case of a confining potential that has spherical symmetry and natural frequency $\omega_{CM}$, the above results can be applied taking $\alpha=\alpha_{z}= \alpha_{\rho}=\sqrt{\pi\hbar/M_T\omega_{CM}}$. Alternatively, we can write
\begin{equation}
 \Phi_{n_{CM},l_{CM},m_{CM}}({\bf R}) =R_{n_{CM},l}(R) Y_{lm}(\theta,\phi),
\end{equation}
where
\begin{equation}R_{n_{CM},l}(R) = N_{n_{CM}l} (R/\sqrt{2}\alpha)^{l}e^{- R^2/2\alpha^2}L_{n_{CM}}^{(l+{1\over 2})}( R^2/\alpha^2),\quad
N_{n_{CM}l}=\sqrt{\sqrt{\frac{1}{\pi }}\frac{2^{n_{CM}+2l+2}\;n_{CM}!\;}{\alpha^3 (2n_{CM}+2l+1)!!}}.\label{eq:ss1}\end{equation}

In that case it results convenient to use Eq.~(\ref{eq:sph}), so that
 \begin{equation}
I^{ST}_{CM}(\mathbf{k};\kappa^{CM}_0;\kappa^{CM}_F) = {\mathcal{I}}^{ST}_{CM}(\hbar\omega^2/M_T\omega_{CM}c^2;\kappa^{CM}_0;\kappa^{CM}_F;\ell)Y_{\ell_{CM},m_{CM}}(\hat{\mathbf{k}}). \label{eq:cmYlma}
\end{equation}
The  radial integral is
\begin{equation}
{\mathrm{I}}^{ST}_{CM}(\hbar\omega^2/M_T\omega_{CM}c^2;n_{CM}^\prime l^\prime;\ell) =\int_0^ \infty dR R^2 R_{n_{CM}^\prime,l^\prime}(R)j_{\ell}(kR)R_{n_{CM},l}(R).
\end{equation}
and the angular integrals can be written in terms of 3-j symbols yielding the coefficients $c(l,m,l^\prime,m^\prime;\ell_{CM},m_{CM})$,
$${\mathrm{I}}^{ST}_{CM}(\hbar\omega^2/M_T\omega_{CM}c^2;\kappa^{CM}_0;\kappa^{CM}_F;\ell)=c(l,m,l^\prime,m^\prime;\ell_{CM},m_{CM}){\mathcal{I}}^{ST}_{CM}(\hbar\omega^2/M_T\omega_{CM}c^2;n_{CM}^\prime l^\prime;\ell)$$

\end{document}